\documentclass[letterpaper]{article} 
\usepackage{aaai25}  
\usepackage{times}  
\usepackage{helvet}  
\usepackage{courier}  
\usepackage[hyphens]{url}  
\usepackage{graphicx} 
\usepackage{natbib}  
\usepackage{caption} 
\usepackage{tabularx} 
\usepackage{float}
\usepackage{algorithm}
\usepackage{algorithmic}
\usepackage{soul}
\usepackage[most]{tcolorbox}
\usepackage{fancyvrb}
\usepackage{xcolor}

\usepackage{newfloat}
\usepackage{listings}
\DeclareCaptionStyle{ruled}{labelfont=normalfont,labelsep=colon,strut=off} 
\floatstyle{ruled}
\newfloat{listing}{tb}{lst}{}
\floatname{listing}{Listing}
\usepackage{multirow}
\usepackage{subcaption}

\newcommand{\vi}[1]{\text{VI}_{\text{#1}}}

\title{Street-Level AI: Are Large Language Models Ready for Real-World Judgments?}
\def\showauthors@on{T}
\author {
    Gaurab Pokharel\textsuperscript{\rm 1},
    Shafkat Farabi\textsuperscript{\rm 1},
    Patrick J. Fowler\textsuperscript{\rm 2}
    Sanmay Das\textsuperscript{\rm 1}
}
\affiliations {
    \textsuperscript{\rm 1}Virginia Tech\\
    \textsuperscript{\rm 2}Washington University in St. Louis\\
    gaurab@vt.edu, mfarabi@vt.edu, pjfowler@wustl.edu, sanmay@vt.edu
}

\usepackage{bibentry}

\begin{document}

\maketitle

\begin{abstract}
    A surge of recent work explores the ethical and societal implications of large-scale AI models that make ``moral'' judgments. Much of this literature focuses either on alignment with human judgments through various thought experiments or on the group fairness implications of AI judgments. However, the most immediate and likely use of AI is to help or fully replace the so-called street-level bureaucrats, the individuals deciding to allocate scarce social resources or approve benefits. 
    There is a rich history underlying how principles of local justice determine how society decides on prioritization mechanisms in such domains. In this paper, we examine how well LLM judgments align with human judgments, as well as with socially and politically determined vulnerability scoring systems currently used in the domain of homelessness resource allocation. Crucially, we use real data on those needing services (maintaining strict confidentiality by only using local large models) to perform our analyses. We find that LLM prioritizations are extremely inconsistent in several ways: internally on different runs, between different LLMs, and between LLMs and the vulnerability scoring systems. At the same time, LLMs demonstrate qualitative consistency with lay human judgments in pairwise testing. Findings call into question the readiness of current generation AI systems for naive integration in high-stakes societal decision-making.
\end{abstract}

%

\section{Introduction}

Large language models (LLMs) have captured public attention and have been broadly touted for their ability to reduce and streamline human work. A question of particular interest of late has been how the moral judgments of LLMs compare with those of humans. This has been engaged in the context of philosophical moral dilemmas \cite{Kim_Kwon_Vecchietti_Oh_Cha_2025, DecodingEthics,rathje2024learning,Dillion_Mondal_Tandon_Gray_2025}, organ transplantation \cite{Murray_et_al_2025,hasjim_et_al_2024}, and fair division \cite{Hosseini_Khanna_2025}, among other domains. With the exception of organ transplantation, most research attends to general conceptions of ethical judgment or definitions of fairness from the fair division community. The central questions of these lines of research have been in trying to understand which general theories of ethics or definitions of fairness LLMs appear to satisfy or whether they replicate certain aspects of human behavior.

However, LLMs are likely to see rapid uptake in domains that involve the allocation of scarce societal resources. The stakes will be more immediate and higher than in toy examples or constructed moral dilemmas, given the scale enabled by AI and the potential number of people affected by such decision-making in high-stakes domains like homelessness services, post-disaster medical triage, and organ transplantation. By analogy to ``vibe coding,'' those working in these domains may feel pressure to use AI to make ``vibe prioritization'' decisions.

Such social and health service settings share key features uniquely relevant for adapting AI applications. First, current delivery reflects \emph{local justice} principles and prioritization practices that emerged to fit the specific contexts over time \cite{elster1992local}. Scarcity requires the adoption of agreed upon rules of who should get what, when, and how.   For example, homelessness services -- the focus of the present study -- currently use a ``vulnerability first'' prioritization, where households deemed to be at greatest risk receive highest priority for resource allocation \cite{kube2023fair}. Medical triage, by contrast, often prioritizes based on maximum expected improvement from receiving the resource. Prioritization frameworks are not inherent and rather emerge from complex social and political processes at play in many countries' social safety net contexts. For homelessness, \emph{implementation} involves the use of point systems based on questionnaire responses administered to at-risk households upon contact with the system (often in so-called ``coordinated entry'' settings). Point systems are what we refer to as \emph{bureaucratic} scoring systems frequently observed in other domains, like public housing and prioritization for liver transplant \cite{johnson2022bureaucratic,been2018allocation,cholongitas2010prioritization}.

Second, actual allocation decisions occur in the province of \emph{street-level bureaucrats} -- civil servants and caseworkers allowed to use their experience to exercise discretion in decision making \cite{Lipsky_2010}. Bureaucratic scoring systems often inform, but do not determine, decisions. 
Bureaucrats themselves bring considerable domain knowledge and expertise from interacting directly with those who ultimately face the outcomes of decisions. Given the embedding within local contexts, street-level judgments likely differ from those of lay humans. Importantly, it is these frontline workers who are likely to be replaced or supplemented by LLMs or ``vibe prioritization.'' 

It is crucial to understand how LLMs would make decisions in such domains and to be as close to reality as possible. Prior work on organ transplantation appears to come closest to doing so. \citet{Murray_et_al_2025} examine the abilities of LLMs to assess medical compatibility in kidney transplantation, as well as the group fairness implications of LLM prioritization. Likewise, \citet{hasjim_et_al_2024} investigate the abilities of LLMs to predict medical benefits and risks of liver transplantation. Yet, neither paper compares LLM prioritization with an established human system of prioritization.

This is the gap that we seek to fill. By systematically querying LLMs, we show that they behave in ways that are qualitatively similar to \emph{lay} humans, as evidenced by a different set of pairwise comparison experiments. Moreover, using a novel dataset of households experiencing homelessness in a major metro area, we compare LLM prioritization with two different standard bureaucratic prioritization methods and assess how well each method predicts the actual allocation decisions of homeless service providers. In contrast with their similarities to lay humans in the simulated task, we find that LLMs are (1) internally inconsistent in the rankings they generate across different runs; (2) inconsistent with the rankings generated by bureaucratic ranking systems; and (3) worse at predicting the allocation decisions of caseworkers (although the scoring systems themselves are not great at this). Taken together, the results seriously question the replacement of street-level bureaucrats with AI in high-stakes societal allocation domains and emphasize the importance of understanding the behavior of front-line workers interacting with the uncertainties inherent in homeless service provision.

\section{Background}

In just the past few years, public interest in LLMs has surged remarkably, and researchers have explored their potential to make moral judgments and guide fair resource allocation. In particular, abstract allocation tasks -- where agents distribute indivisible goods and, optionally, monetary compensation among individuals with heterogeneous valuations -- have assessed LLM behavior against foundational fairness axioms such as equitability and envy‐freeness. \citet{Hosseini_Khanna_2025} show that although LLMs can sometimes satisfy single fairness criteria, their choices often diverge from human distributional preferences and are sensitive to prompt phrasing and persona assignments. Complementing these findings, \citet{Kim_Kwon_Vecchietti_Oh_Cha_2025} document substantial variance in moral reasoning when models adopt different sociodemographic personas, with politically charged framings exacerbating bias and polarization. At the same time, LLMs hold the promise of articulating trade-offs and justifying their decisions in rich natural language rationales -- fueling an emerging concern: do these ``toy’’ successes hold up under the pressures of real-world stakes?

Such stylized experiments, whether dividing cookies, dollars, or chores, inevitably abstract away the contextual nuances and ethical complexities of high‐stakes domains. Yet, the public increasingly perceives LLM outputs as possessing moral expertise comparable to or exceeding that of professional ethicists \cite{Dillion_Mondal_Tandon_Gray_2025}. Misplaced confidence underscores a key limitation: high‐stakes domains demand more than metaphorical toy examples and call for careful evaluation in concrete, socially consequential contexts.

To move beyond the question of ``which pie slice feels fairest’’ and toward ``who receives life-altering resources first,’’ we must ground our inquiry in a domain defined by: first, the existence of standardized triage tools; second, access to rich real-world data; and third, clear human benchmarks against which to judge model performance. Homelessness resource allocation is one such domain. Communities across North America establish protocols -- commonly referred to as coordinated entry systems -- to assess household vulnerability and to prioritize scarce homelessness services for the neediest.

Operationalizing this vulnerability-forward framework, the Vulnerability Index-Service Prioritization Decision Assistance Tool (VI-SPDAT) suite — the core VI-SPDAT \cite{orgcode2015vispdat} and its derivatives, the VI-F-SPDAT for families \cite{orgcode2015vifspdat} and the TAY-VI-SPDAT for transition-aged youth \cite{orgcode2015tayvispdat} - consists of brief, self-reported intake questionnaires covering domains, such as housing history, health status, and social support. Responses are aggregated through a predefined rubric into a composite acuity score, allowing minimally trained assessors to triage clients rapidly. To date, these tools have been deployed in over one thousand communities in North America \cite{orgcode2015vispdat}. Despite widespread use, the VI-SPDAT faces criticism for introducing biases against certain subpopulations and for oversimplifying complex needs \cite{VISPDAT_validity,cronley2022invisible,shinn2022allocating}.

\subsubsection{Bureaucrats in Social Services: } The responsibility for helping navigate this intricate triage process is typically entrusted to \emph{street‐level bureaucrats}, who wield significant discretion in policy interpretation and implementation \cite{Lipsky_2010}. Bureaucrats decipher policies and allocate homeless resources under tight budgets, standardization pressures, and their own professional judgment. \citet{Pokharel_Das_Fowler_2024} show that, at a time when intervention assignments were not formulaic, beyond rule-based assignments, caseworkers exercise discretion that sometimes favors less vulnerable households but delivers larger marginal gains when more intensive services are provided, highlighting how experiential knowledge complements standardized scores. 

Efforts to introduce caseworkers to algorithmic decision support tools are not new and have met resistance in the past. In child welfare, bureaucrats reject recommendations deemed contextually inappropriate despite institutional conformity pressures \cite{childWelfare2022}. Furthermore, front-line workers and unhoused individuals alike voice concerns about biases in AI-driven homelessness services, underscoring tensions between algorithmic objectives and individual needs \cite{homelessnesPerspective2023}.

Illustrating these dynamics in homelessness services, \citet{kube2022just} enlisted 458 Mechanical Turk participants to simulate street-level allocation decisions by completing ten pairwise comparisons of homeless household profiles drawn from St. Louis’s Homeless Management Information System (HMIS) -- randomizing subjects to view either only demographic and service-request data or that same data augmented with Bayesian Additive Regression Trees derived low/medium/high risk scores of reentering homelessness within two years. Choices clustered into two distinct prioritization styles -- ``vulnerability-oriented’’ (favoring households with higher reentry risk) and ``outcome-oriented’’ (favoring those with lower risk) -- and prior exposure to algorithmic predictions both nudged undecided participants toward outcome-oriented allocations and amplified each individual’s intrinsic decision-making style. These results highlight how even relatively simple risk scores can subtly but systematically reshape frontline judgments, underscoring the complexity of introducing algorithmic support into real-world resource-allocation contexts.

At the same time, LLMs have been enthusiastically deployed in other social service domains, often without much scrutiny \cite{chatbot_social_services,Taylor_2024,real_dangers_AI}. Empirical studies expose their brittleness and fairness vulnerabilities: small input tweaks can drastically alter clinical risk recommendations \cite{acharya_et_al_2024}, shift demographic outcomes under Borda‐based rankings \cite{Murray_et_al_2025}, or overlook subtle but critical factors in AI‐driven committee decisions \cite{hasjim_et_al_2024}. Theoretical work also warns about interpretability challenges and the potential for rapid, catastrophic errors in the absence of human oversight \cite{Conitzer_2024}.

The documented vulnerabilities, alongside the critical and nuanced nature of the problem, beg the question: Should LLMs be considered in the context of homelessness resource allocation at all? Our analysis specifically examines the alignment of LLM judgments with both human caseworker assessments and established vulnerability scoring systems in homelessness resource allocation. Leveraging real-world intake data, we investigate LLM internal consistency, agreement across model variants, and compatibility with existing social and political prioritization mechanisms, insights that are crucial for assessing the practical viability of incorporating LLMs into high-stakes societal decision-making contexts.

\begin{figure*}[h]
    \centering
    \includegraphics[width=\textwidth]{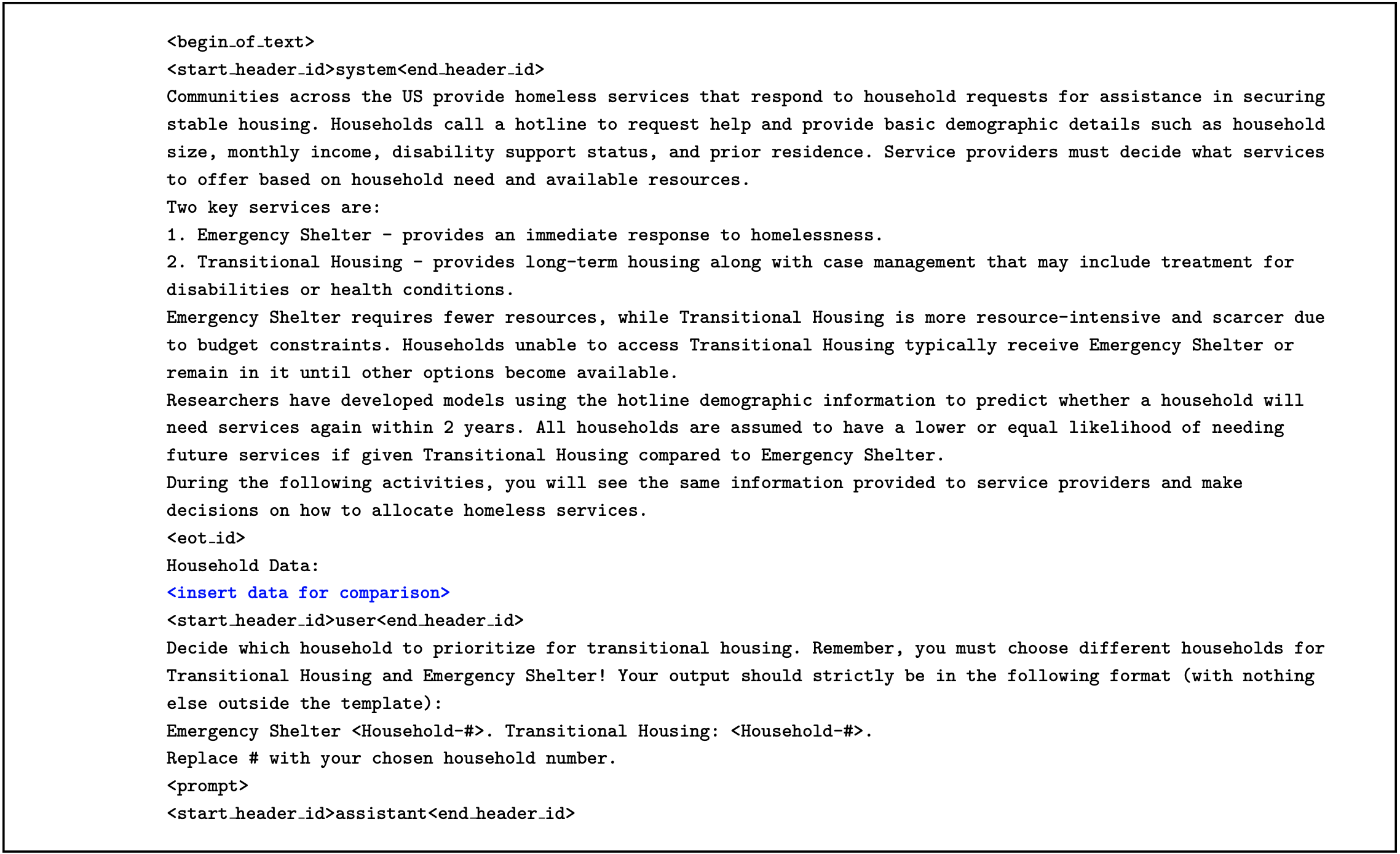}
    \caption{The `base prompt' used in both the pairwise comparison and ranking tasks. Depending on the experiment type, we replace the blue placeholder text with either the tabular data (for the pairwise comparison) or the questionnaire responses (for the ranking task).}
    \label{fig:sample_prompt}
\end{figure*}

\section{Methods}

To assess how well LLMs align with both human judgments and established vulnerability scoring systems in homelessness resource allocation, we design two complementary experimental tasks:

\begin{itemize}
    \item [1] \emph{Pairwise Comparisons}: We replicate an experiment by \citet{kube2022just} who used 10 pairs of household data drawn from St. Louis homelessness services records (including features on demographics, income, disability, service requests, and reentry to homelessness risk scores). The main purpose is to investigate whether the LLM being tested is more ``outcome-oriented'' or ``vulnerability-oriented'' in deciding which household to allocate a more intensive intervention and how the orientations align with lay humans. We limit our experiments to the same 10 pairs of household data \citet{kube2022just} used in order to accurately compare the LLM allignments with humans. 
    \item [2] \emph{Ranking Task}: In this experiment, we use data on homeless service vulnerability assessments collected from 2021 through 2024 in St. Louis and elicit pairwise prioritization comparisons from several LLMs to create a complete ranked list of households for prioritization. We then compare the LLM rankings with those obtained from standard rankings actually used in the real-world prioritization scheme. 
\end{itemize}

\noindent In both tasks, we measure internal consistency (variance across runs), inter‐model agreement, and alignment with established vulnerability scores using Spearman’s $\rho$ \cite{Spearman_1904}, a non-parametric statistic that measures the strength and direction of association between two variables by comparing the ranked order of the values. We now explain the two tasks in detail. 

\subsection{Pairwise Comparisons}
\label{sec:Pairwise}

\citet{kube2022just} drew on St. Louis’s Homeless Management Information System (HMIS) to create ten hypothetical pairs of tabular data of households. These data tables are already public and published on the web, thus we are also able to use cloud-based proprietary models for evaluation. The data contained information on demographics (number of adults and children, monthly income, disability status), service requests, and a precomputed risk score (low/medium/high probability of reentering homelessness within two years). In that experiment, human subjects mimicked street-level bureaucrats who were told that two types of assistance were available: 1) \textbf{Transitional Housing (TH)}, which is a longer-term and more intensive support program, reserved for households with greater needs; and, 2) \textbf{Emergency Shelter (ES)} -- a less intensive option that provides a place to stay overnight. The lay humans were then presented with features of household pairs and asked to prioritize one for TH, with the understanding that the other would receive ES. 

We run a similar set of experiments with LLMs. We test four information conditions:
\begin{itemize}
    \item \textbf{No Prediction:} The model sees only the household data table (no reentry to homelessness risk estimates).
    \item \textbf{Only Prediction:} The model sees only the outcome prediction (no data table).
    \item \textbf{Prediction First:} The model first generates its own outcome prediction from the data tables provided and \emph{then} makes the allocation decision.
    \item \textbf{Shared Prediction:} The model sees both the household data table \emph{and} a pre-computed ``outcome prediction'' (the probability -- labeled low/medium/high --that the household will reenter homelessness services within two years, conditional on receiving each intervention).
\end{itemize}

\noindent These four setups allow us to identify how access to risk estimates affects the LLM’s choices. Then, for each of the ten household pairs, we issue ten independent prompts per condition for every LLM that we test. Each response is labeled
\emph{outcome-oriented} (favoring the lower-risk household) or \emph{vulnerability-oriented} (favoring the higher-need household). We then compute an \emph{outcome score} for each pair:

\begin{equation}
    \mathrm{Outcome\ Score}
    \;=\;
    \left\lceil
      \frac{\text{\# of outcome decisions}}{\text{total decisions}}
      \times 100
    \right\rceil
    \label{fig:eq}
\end{equation}

\noindent An outcome score of 0 means that all ten runs were vulnerability‐oriented; a score of 100 means that all were outcome‐oriented. The exact prompt and details on the LLMs used are provided in Section \ref{subsection:prompt_details}.

\subsection{Ranking Task}

The second set of experiments evaluates how LLMs perform in the core bureaucratic task of coordinated entry -- that is, producing a complete ordering of a given set of households by relative vulnerability. In practice, such rankings drive the allocation of scarce resources according to socially and politically defined priorities. We compare the LLM-generated rankings with those produced by established assessment instruments: the VI-SPDAT for single adults, families, and youth (hereafter denoted $\vi{SA}$, $\vi{F}$, and $\vi{Y}$ respectively) \cite{orgcode2015vispdat,orgcode2015vifspdat,orgcode2015tayvispdat}, and the Risk/Medical Frailty Score (RMFS), another tool used in St. Louis homeless services \cite{fefer2022}. Our dataset contains, for each household, the VI-SPDAT acuity score, the RMFS frailty score, and the full set of raw questionnaire responses. These questions aim to assess vulnerability. Examples include \emph{How long has it been since you lived in permanent, stable housing?}, and \emph{Where do you sleep most frequently?} There are in total 35 questions in $\vi{SA}$, 41 questions in $\vi{Y}$,  54 questions in $\vi{F}$, and 25 in RMFS. The full set of questions used to collect the data is included in Appendix \ref{appendix:ques} and \ref{appendix:ques_rmfs}. Table \ref{table:ranking_num_data} documents the number of available assessments. We refer to these scores collectively as bureaucratic rankings. 

It is widely acknowledged that asking an LLM to rank directly yields unreliable global orderings.  We follow the recommendation of \citet{qin-etal-2024-large} and instead  use a methodology that constructs the ranking from pairwise comparisons using Rank Centrality \citep{rank_centrality}. 
We start by comparing each pair of households with the LLM and recording if $i$ is preferred to $j$. For each ordered pair $(i,j)$, we compute the raw ``win fraction'' and treat that as the weight on the edge between $i$ and $j$. To convert raw weights into transition probabilities for a random walk, we \emph{normalize} each household's outgoing edges so that they sum to one by simply dividing each edge-weight by the total of that node's outgoing weights. This ensures each node’s edges form a valid probability distribution. 

We then interpret the normalized edge weights as transition probabilities of a Markov chain and compute the stationary distribution of this Markov chain to recover a global score for each household. The rank centrality algorithm is closely connected to the \emph{Bradley–Terry–Luce (BTL)} model \cite{bradley1952rank,luce1959individual}, which assumes that each household $i$ has a positive ``strength'' $\theta_i$, and that in any single comparison the probability $i$ beats $j$ is $\frac{\theta_i}{\theta_i + \theta_j}$. Under this model, the stationary distribution of the normalized comparison graph converges to the true normalized strengths, and it can be shown that only on the order of $\theta(N \log N)$ independent comparisons are required to recover accurate rankings with high probability \citep{rank_centrality}. Since exhaustively comparing all $\frac12N(N-1)$ pairs is infeasible, we instead include each (unordered) pair independently with probability $0.4$, yielding about $0.2\,N(N-1)$ comparisons—comfortably above the $\theta(N \log N)$  threshold—while keeping compute and LLM calls tractable (see Table~\ref{table:ranking_num_data}). Importantly, throughout this entire process, we never send data to the cloud. All computations and LLM inferences are performed locally, ensuring strict data privacy and confidentiality. Our decision to avoid cloud-based LLM calls constrains us to utilize only models locally available, leading us to employ different sets of models for the pairwise comparison task and the ranking task.

\begin{figure*}[h]
    \centering
    \includegraphics[width=0.55\textwidth]
    {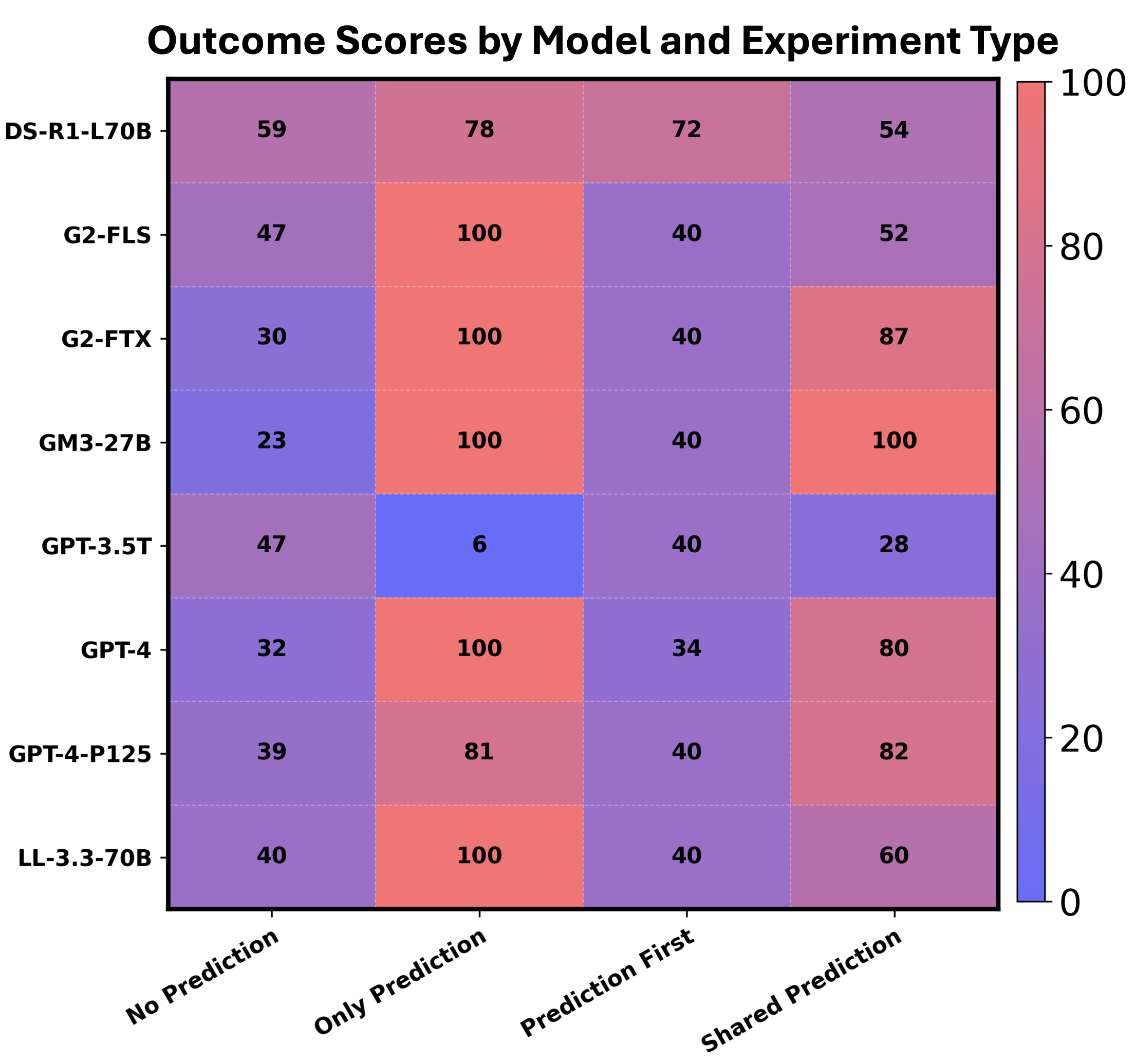}
    \caption{Outcome-oriented scores from our pairwise comparison task, broken down by model and information configuration. In the heatmap, red denotes higher scores (more outcome-oriented) and blue denotes lower scores (more vulnerability-oriented). These results reveal substantial variability in LLM behavior across different models—and show that, overall, LLMs perform similarly to non-expert humans when making prioritization judgments.
 }
    \label{fig:model_level_results}
\end{figure*}

Each time we sample an edge, the LLM decides which household ``wins'', that is, the model judges it more vulnerable. Thus, the sampled edge accrues a higher transition probability in the rank-centrality graph. Consequently, when we compute the stationary distribution of this Markov chain, those households deemed most vulnerable by the LLM naturally receive the highest global scores in the final aggregated ranking.

\subsection{LLMs and Prompting}
\label{subsection:prompt_details}

The details of the LLMs that we use (and their aliases) are detailed in Table \ref{tab:model_nicknames}.


\begin{table}[ht]
\resizebox{\columnwidth}{!}{%
\begin{tabular}{|l|l|}
\hline

\multicolumn{2}{|c|}{\textbf{Pairwise Comparison}} \\ \hline
\textbf{Full Model Name} & \textbf{Alias} \\ \hline
DeepSeek-R1-Distill-Llama-70B  \cite{deepseekai2025deepseekr1incentivizingreasoningcapability} & DS-R1-L70B \\
Llama-3.3-70B-Instruct  \cite{llama} & LL-3.3-70B \\
Gemini-2.0-flash-thinking-exp \cite{gemini-2.0-flash-thinking-exp} & G2-FTX \\
Gemini 2.0 flash \cite{gemini-2.0-flash} & G2-FLS \\
Gemma-3-27b-it \cite{gemma-3-27b-it} & GM3-27B \\
GPT-4 \cite{gpt-4} & GPT-4 \\
GPT-3.5-turbo \cite{gpt-3.5-turbo} & GPT-3.5T \\
GPT-4.0125-preview \cite{gpt-4-0125-preview} & GPT-4-P125 \\ \hline
\multicolumn{2}{|c|}{\textbf{Ranking Task}} \\ \hline
Llama-3-8B-Instruct \cite{llama3modelcard} & LL-3-8B \\
deepseek-llm-7b-chat  \cite{deepseek-llm} & DS-7B \\ \hline
\end{tabular}%
}
\caption{The LLMs used in our experiments and their shorthand aliases. Models are grouped by task.}
\label{tab:model_nicknames}
\end{table}

\subsubsection{Pairwise Comparisons: } For the pairwise comparison task, because the data is publicly available in \citet{kube2022just} structured as \emph{blocks} numbered one through ten, we evaluated both open-source and proprietary LLMs. The open source models included DS-R1-L70B and LL-3.3-70B, while the proprietary models comprised G2-FTX, G2-FLS, GM3-27, GPT-4, GPT-3.5T, and GPT 4-P125. We take the data from \citet{kube2022just} and convert them into a json-like string and use a base prompt (see Figure \ref{fig:sample_prompt}) to replace the households data for each pair, such that we are working with the same prompt for every pairwise comparison. 

\subsubsection{Ranking Task: } As mentioned above, we sample approximately 40\% of the undirected edges for the rank centrality algorithm. In order to derive edge weights, we need the outcomes of pairwise comparisons. We reuse the same base prompt from the pairwise experiments, but replace the data now with the responses from each household's raw questionnaire. Given the proprietary nature of our questionnaire data and to ensure that information never leaves our secure environment, we only use open-source LLMs that can run locally. Considering the computational requirements for the large number of prompts necessary, we chose two 7-billion-parameter models -- LL-3-8B  and DS-7B -- that come from different development ecosystems (the former rooted in Western open-source communities, the latter from Chinese academic collaborations). This allows us to explore whether varying design philosophies or cultural contexts affect vulnerability judgments. For each LLM, we execute the full rank-centrality pipeline twice, producing two independent ranked lists per model to quantify the consistency of their rankings. 

All local models were run on a cluster compute node containing two 24GB VRAM Nvidia A30 GPUs using vLLM's api-server's OpenAI compatible endpoints. Both models were deployed using publicly available pre-trained weights from the Hugging Face Hub, and no additional fine-tuning was performed. We opted not to fine-tune the models because our core research question is: Which types of households do  current state-of-the-art LLMs, without any form of fine-tuning, consider more vulnerable / needing a more intensive intervention? Evaluating ``vanilla'' models tells us how well they transfer to a novel, socially important task without bespoke adaptation, which mirrors many real-world settings where practitioners lack labeled data, compute budgets, or permission to retrain proprietary APIs. This is the sense in which we can think of this ``vibe prioritization'' as an analogy to ``vibe coding,'' where humans ask AI models to produce code with minimal intervention or oversight. We note that we access the proprietary models via API tokens, but follow the same experimental pipeline as our local models.

For most models, we use regular expressions to parse the LLM outputs. DS-7B, however, produces overly verbose responses, so we route its output through a secondary LLM (G2-FLS) to distill the household selections. On the rare occasions when the models failed to yield a coherent choice, we re-prompt until the LLM unambiguously identifies which household should receive the more intensive intervention. 

\section{Results}
\label{section:results}

\begin{figure*}[h]
  \centering
  \begin{subfigure}{0.495\textwidth}
    \includegraphics[width=\linewidth]{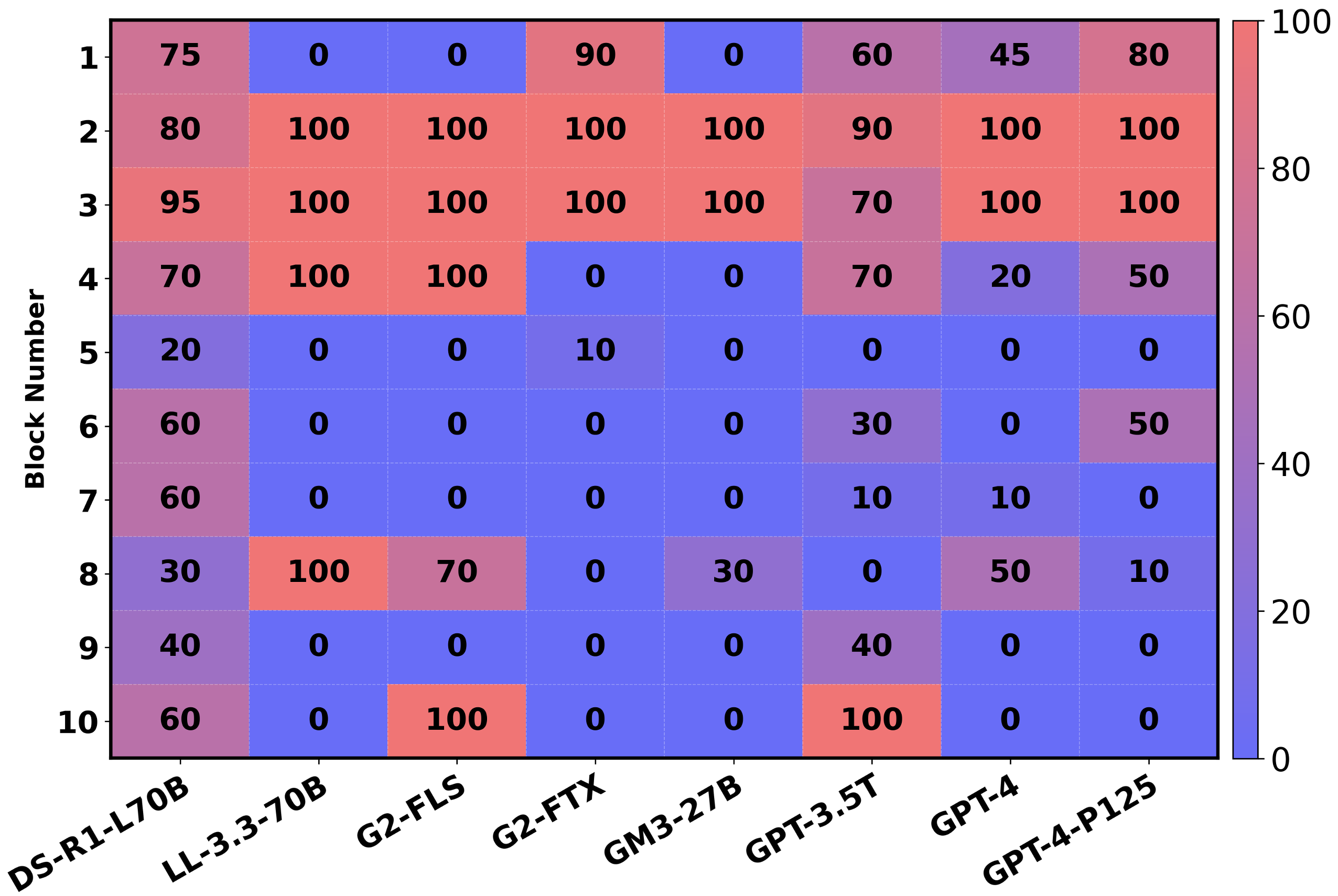}
    \caption{No Prediction}
    \label{fig:block_blind}
  \end{subfigure}
  \begin{subfigure}{0.495\textwidth}
    \includegraphics[width=\linewidth]{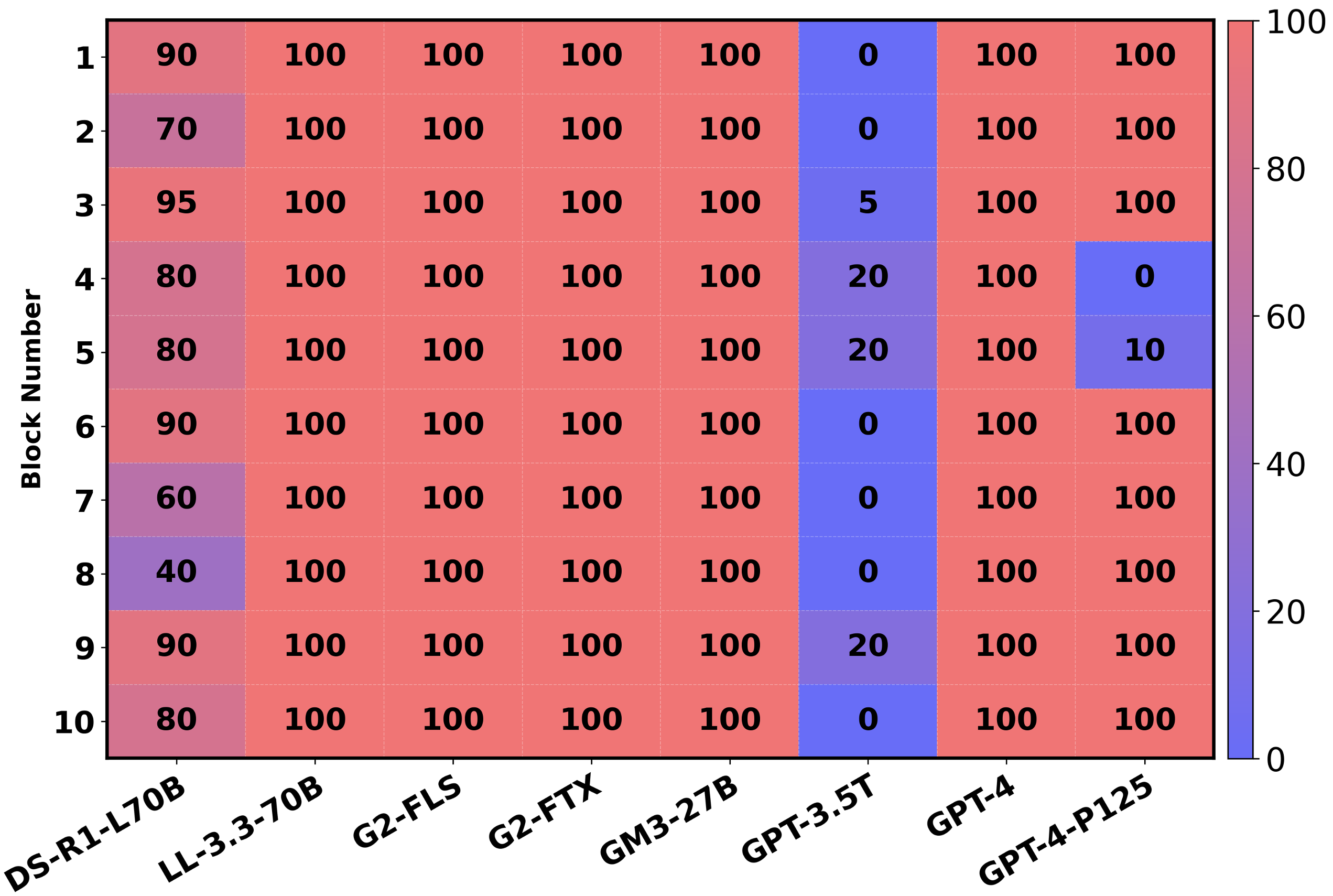}
    \caption{Only Prediction}
    \label{fig:block_vision_only}
  \end{subfigure}
  \vskip1.5\baselineskip
  \begin{subfigure}{0.495\textwidth}
    \includegraphics[width=\linewidth]{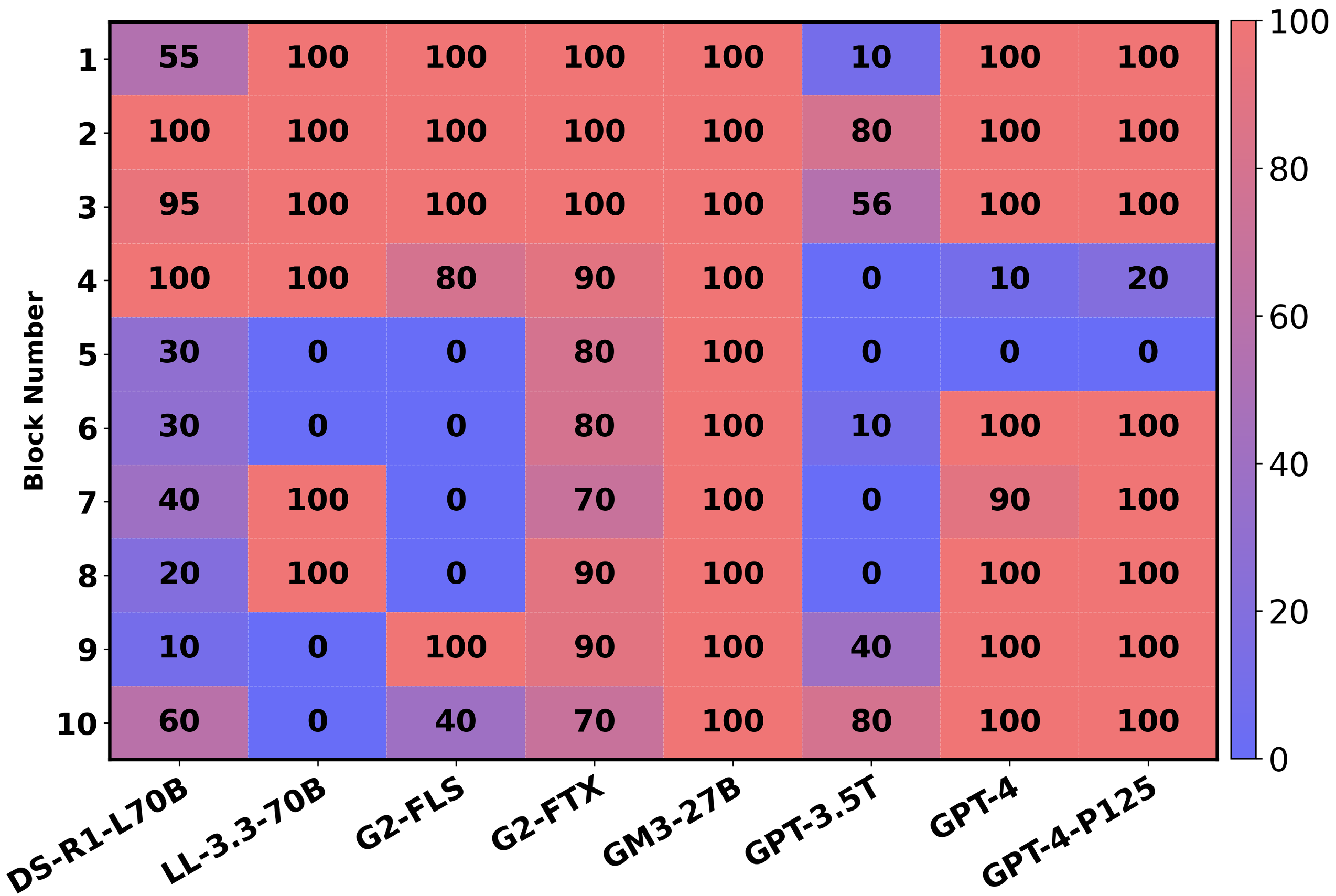}
    \caption{Shared Prediction}
    \label{fig:block_shared_vision}
  \end{subfigure}
  \begin{subfigure}{0.495\textwidth}
    \includegraphics[width=\linewidth]{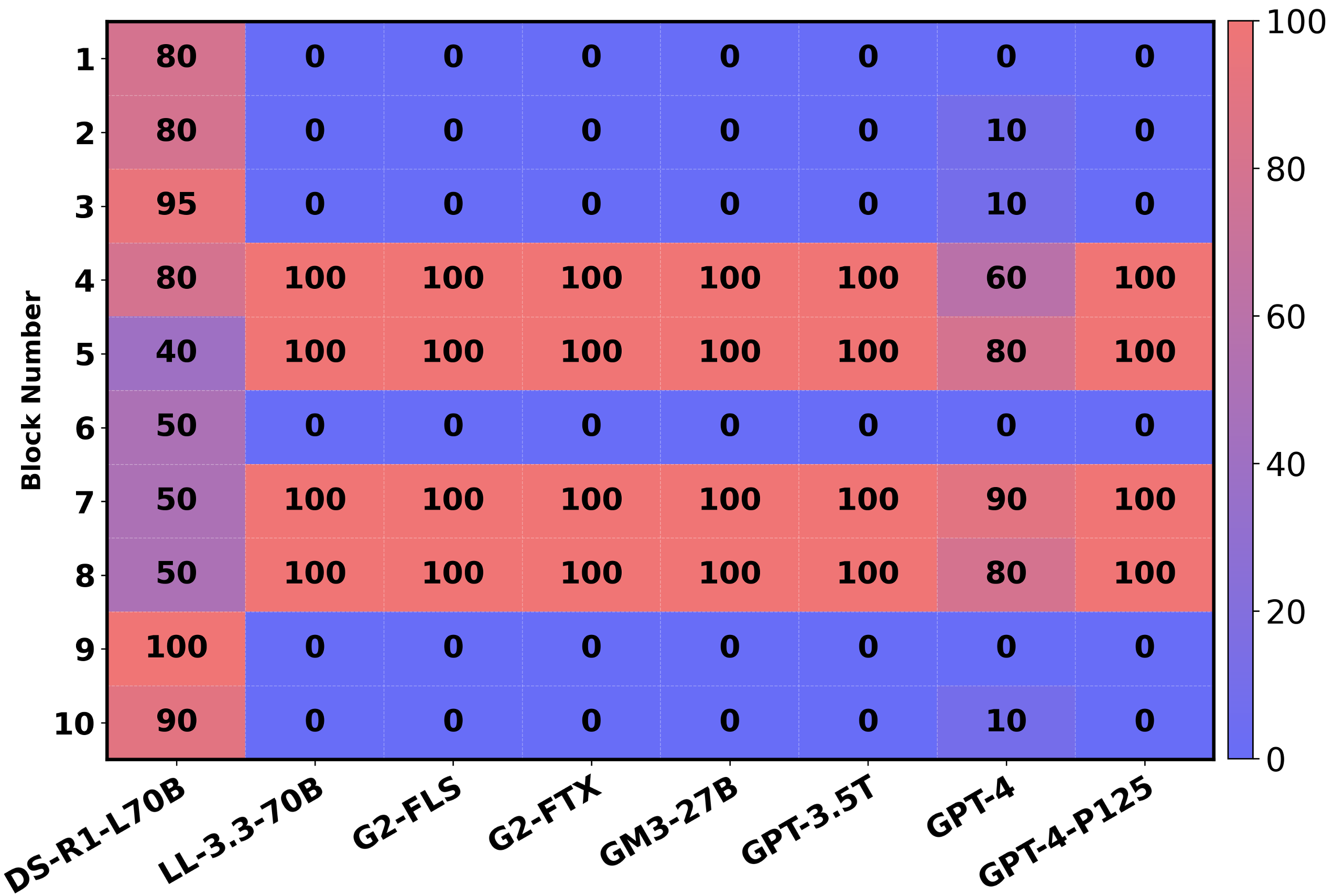}
    \caption{Prediction First}
    \label{fig:block_vision_first}
  \end{subfigure}
  \caption{Pairwise comparison results under four conditions: no predictions (Fig. \ref{fig:block_blind}); only outcome predictions (Fig. \ref{fig:block_vision_only}); shared prediction (Fig. \ref{fig:block_shared_vision}); and prediction first (Fig. \ref{fig:block_vision_first}). Each block corresponds to a household. Scores in each cell are calculated using Equation \ref{fig:eq} from outcome decisions averaged across 10 different runs. These results indicate that LLM decision inconsistencies do not occur within the block level, but instead across blocks.}
  \label{fig:block_level_results}
\end{figure*}


\subsection{Pairwise Comparisons}
\label{subsec:pariwise_results}

We repeatedly prompted each LLM in the pairwise comparison experiments until it produced a coherent, categorizable response.\footnote{DS-R1-L70B often diverged into tangents, and GPT-4 initially refused on grounds of unqualification; nonetheless, all models ultimately yielded categorizable outputs after repeated prompting.} Our main goal here is to compare LLM prioritization behavior with that observed in lay humans on carefully constructed data. This is similar in spirit to much of the ongoing work on LLMs making moral judgments in different domains \cite{Kim_Kwon_Vecchietti_Oh_Cha_2025, DecodingEthics,Dillion_Mondal_Tandon_Gray_2025,nazi2024large}. We calculate the outcome scores of the decisions rendered by different LLMs under different configurations and average them over household pairs. Figure \ref{fig:model_level_results} presents the outcome scores of the results, where each column corresponds to the different experiment types and each row represents the LLMs tested for the task. Recall that scores range from 0 (completely vulnerability oriented, colored blue) to 100 (completely outcome oriented, colored red).

 \paragraph{No Prediction.} When no outcome information is provided (only tabular data), LLMs are, on average, inconsistent in terms of whether they present as vulnerability- or outcome- oriented in decision-making. This is not due to inconsistency across runs within a specific pairwise comparison, but instead due to making different decisions across pairs (see Figure \ref{fig:block_level_results}). This is similar to the behavior Kube et al observe in humans, indicating that LLMs may have similar difficulty in assessing vulnerability or outcomes from the raw data.

 \paragraph{Only Prediction.} When \emph{only} outcome information is provided, similar to humans, each LLM is highly consistent in manifesting either a vulnerability or an outcome orientation. Notably, GPT-3.5T is the only LLM that is vulnerability-oriented,  with an outcome score of 6. 

 \paragraph{Prediction First.} Kube et al. provide evidence that prior exposure to quantitatively predicting outcomes can alter the orientation of some (about one third) of humans from vulnerability to outcome. For the LLMs we tested, the outcome orientation is not substantively different from the No Prediction condition, indicating that there is no such ``framing'' effect here.

 \paragraph{Shared Prediction.} This experiment demonstrates substantial heterogeneity between different LLMs. In human behavior, when comparing No Prediction to Shared Prediction, Kube et al. note that the addition of outcome information allows each human to manifest their ``true'' type (equivalent, in our setting, to whether or not they appear vulnerability- or outcome-oriented in the Prediction Only task). In fact, we see a similar movement of the LLMs. Each moves towards their Prediction Only score, with the exception of DS-R1-L70B. GS-FLS is somewhat limited in its movement. 

 Together, the results indicate substantial heterogeneity in LLM behavior across different models. They also show that LLMs in general are not dissimilar from lay humans (with no specific subject matter expertise) in making judgments on the problem of prioritization.


\begin{table*}[]
\resizebox{\textwidth}{!}{%
\begin{tabular}{lclllc}
\multicolumn{2}{c}{LLaMA Ranking 1 on $\vi{SA}$} &  & \multicolumn{1}{c}{} & \multicolumn{2}{c}{LLaMA Ranking 2 on $\vi{SA}$} \\ \cline{1-2} \cline{5-6} 
\multicolumn{1}{|l|}{\textbf{Feature}} & \multicolumn{1}{c|}{\textbf{Coefficient}} &  & \multicolumn{1}{l|}{} & \multicolumn{1}{l|}{\textbf{Feature}} & \multicolumn{1}{c|}{\textbf{Coefficient}} \\ \cline{1-2} \cline{5-6} 
\multicolumn{1}{|l|}{Current living arrangement: Shelter/Outdoors } & \multicolumn{1}{c|}{{\color[HTML]{FE0000} -0.3796}} &  & \multicolumn{1}{l|}{} & \multicolumn{1}{l|}{{\color[HTML]{3531FF} Current living arrangement: No answer}} & \multicolumn{1}{c|}{{\color[HTML]{FE0000} -0.0632}} \\
\multicolumn{1}{|l|}{{\color[HTML]{3531FF} Current living arrangement: Car}} & \multicolumn{1}{c|}{{\color[HTML]{009901} 0.0216}} &  & \multicolumn{1}{l|}{} & \multicolumn{1}{l|}{Current living arrangement: Job corp} & \multicolumn{1}{c|}{{\color[HTML]{FE0000} -0.0290}} \\
\multicolumn{1}{|l|}{{\color[HTML]{3531FF} Current living arrangement: No answer}} & \multicolumn{1}{c|}{{\color[HTML]{009901} 0.0517}} &  & \multicolumn{1}{l|}{} & \multicolumn{1}{l|}{{\color[HTML]{3531FF} Current living arrangement: Car}} & \multicolumn{1}{c|}{{\color[HTML]{FE0000} -0.0283}} \\
\multicolumn{1}{|l|}{Attempts of self-harm? No answer} & \multicolumn{1}{c|}{{\color[HTML]{009901} 0.1025}} &  & \multicolumn{1}{l|}{} & \multicolumn{1}{l|}{Current living arrangement: Hotel} & \multicolumn{1}{c|}{{\color[HTML]{FE0000} -0.0236}} \\
\multicolumn{1}{|l|}{Number of past incarcerations $>$ 10} & \multicolumn{1}{c|}{{\color[HTML]{009901} 0.1296}} &  & \multicolumn{1}{l|}{} & \multicolumn{1}{l|}{Current living arrangement: Couch-surfing/Shelter} & \multicolumn{1}{c|}{{\color[HTML]{009901} 0.5033}} \\ \cline{1-2} \cline{5-6} 
 & \multicolumn{1}{l}{} &  &  &  & \multicolumn{1}{l}{} \\
\multicolumn{2}{c}{LLaMA Ranking 1 on $\vi{F}$} &  & \multicolumn{1}{c}{} & \multicolumn{2}{c}{LLaMA Ranking 2 on $\vi{F}$} \\ \cline{1-2} \cline{5-6} 
\multicolumn{1}{|l|}{\textbf{Feature}} & \multicolumn{1}{c|}{\textbf{Coefficient}} &  & \multicolumn{1}{l|}{} & \multicolumn{1}{l|}{\textbf{Feature}} & \multicolumn{1}{c|}{\textbf{Coefficient}} \\ \cline{1-2} \cline{5-6} 
\multicolumn{1}{|l|}{{\color[HTML]{3531FF} Current living arrangement: No answer}} & \multicolumn{1}{c|}{{\color[HTML]{FE0000} -0.0469}} &  & \multicolumn{1}{l|}{} & \multicolumn{1}{l|}{{\color[HTML]{3531FF} Current living arrangement: No answer}} & \multicolumn{1}{c|}{{\color[HTML]{FE0000} -0.0145}} \\
\multicolumn{1}{|l|}{Current living arrangement: Couch-surfing} & \multicolumn{1}{c|}{{\color[HTML]{FE0000} -0.0228}} &  & \multicolumn{1}{l|}{} & \multicolumn{1}{l|}{Expected additional children to join after receiving housing: No answer} & \multicolumn{1}{c|}{{\color[HTML]{009901} 0.0619}} \\
\multicolumn{1}{|l|}{Current living arrangement: Place-to-place} & \multicolumn{1}{c|}{{\color[HTML]{FE0000} -0.0135}} &  & \multicolumn{1}{l|}{} & \multicolumn{1}{l|}{Parents in family $>$ 3} & \multicolumn{1}{c|}{{\color[HTML]{009901} 0.0757}} \\
\multicolumn{1}{|l|}{{\color[HTML]{3531FF} Current living arrangement: Car}} & \multicolumn{1}{c|}{{\color[HTML]{FE0000} -0.0081}} &  & \multicolumn{1}{l|}{} & \multicolumn{1}{l|}{Nights spent incarcerated $>$ 3} & \multicolumn{1}{c|}{{\color[HTML]{009901} 0.1110}} \\
\multicolumn{1}{|l|}{Current living arrangement: Couch-hopping} & \multicolumn{1}{c|}{{\color[HTML]{009901} 0.5707}} &  & \multicolumn{1}{l|}{} & \multicolumn{1}{l|}{{\color[HTML]{3531FF} Current living arrangement: Car}} & \multicolumn{1}{c|}{{\color[HTML]{009901} 0.1511}} \\ \cline{1-2} \cline{5-6} 
 & \multicolumn{1}{l}{} &  &  &  & \multicolumn{1}{l}{} \\
\multicolumn{2}{c}{LLaMA Ranking 1 on $\vi{Y}$} &  & \multicolumn{1}{c}{} & \multicolumn{2}{c}{LLaMA Ranking 2 on $\vi{Y}$} \\ \cline{1-2} \cline{5-6} 
\multicolumn{1}{|l|}{\textbf{Feature}} & \multicolumn{1}{c|}{\textbf{Coefficient}} &  & \multicolumn{1}{l|}{} & \multicolumn{1}{l|}{\textbf{Feature}} & \multicolumn{1}{c|}{\textbf{Coefficient}} \\ \cline{1-2} \cline{5-6} 
\multicolumn{1}{|l|}{Reselling prescribed painkiller? No answer} & \multicolumn{1}{c|}{{\color[HTML]{FE0000} -0.0558}} &  & \multicolumn{1}{l|}{} & \multicolumn{1}{l|}{Did abusive relationship cause homelessness? No answer} & \multicolumn{1}{c|}{{\color[HTML]{FE0000} -0.1244}} \\
\multicolumn{1}{|l|}{Number of ER visits $>$ 10} & \multicolumn{1}{c|}{{\color[HTML]{009901} 0.0929}} &  & \multicolumn{1}{l|}{} & \multicolumn{1}{l|}{Marijuana use before age of 12? No answer} & \multicolumn{1}{c|}{{\color[HTML]{FE0000} -0.0588}} \\
\multicolumn{1}{|l|}{Avoids seeking medical help? No answer} & \multicolumn{1}{c|}{{\color[HTML]{009901} 0.0969}} &  & \multicolumn{1}{l|}{} & \multicolumn{1}{l|}{Chronic health issues? No answer} & \multicolumn{1}{c|}{{\color[HTML]{FE0000} -0.0505}} \\
\multicolumn{1}{|l|}{Current living arrangement: Boyfriend's house} & \multicolumn{1}{c|}{{\color[HTML]{009901} 0.0981}} &  & \multicolumn{1}{l|}{} & \multicolumn{1}{l|}{Time since permanent housing? No answer} & \multicolumn{1}{c|}{{\color[HTML]{009901} 0.0547}} \\
\multicolumn{1}{|l|}{Physical health impacted homelessness: No answer} & \multicolumn{1}{c|}{{\color[HTML]{009901} 0.1353}} &  & \multicolumn{1}{l|}{} & \multicolumn{1}{l|}{Homelessness caused by running away? No answer} & \multicolumn{1}{c|}{{\color[HTML]{009901} 0.1200}} \\ \cline{1-2} \cline{5-6} 
\end{tabular}%
}
\caption{Top 5 most influential question-answer pairs for two independent LLaMA rankings. The low overlap and inconsistent polarity of the few shared features (highlighted in blue) demonstrate the model's instability. Questions are shortened for brevity (see Appendix \ref{appendix:ques} \cite{full_paper} for full versions.)}
\label{fig:top5_featuresllama}
\end{table*}

\subsection{Rankings}
\label{subsec:rankings_results}

Before presenting results for the ranking task, we address how outcome score ties affect score-based rankings, since multiple households may receive the same score. To evaluate the impact of random tie breaking on the overall order, we generate ten tie-broken variants for each assessment tool by permuting tied entries uniformly at random. In all cases, Spearman’s $\rho$ between each pair of variants remained above 0.98, indicating that arbitrary tie resolutions do not significantly alter the global rank structure. To provide a complete statistical picture, 95\% confidence intervals for this and all following analyses are in Table \ref{tab:all_results}, Appendix \ref{Appendix:All_Tables}\footnote{Full paper on arXiv \cite{full_paper}}.\\

\begin{table}[ht]
\resizebox{\columnwidth}{!}{%
\begin{tabular}{|l|l|l|l|l|}
\hline
\textbf{\begin{tabular}[c]{@{}l@{}}Assessment \\ Type\end{tabular}} & \textbf{\begin{tabular}[c]{@{}l@{}}Dataset\\ Name\end{tabular}} & \textbf{\begin{tabular}[c]{@{}l@{}}\# \\ Samples\end{tabular}} & \textbf{\begin{tabular}[c]{@{}l@{}}Min Corr\\ (SPDAT)\end{tabular}} & \textbf{\begin{tabular}[c]{@{}l@{}}Min Corr\\ (RMFS)\end{tabular}} \\ \hline
\textbf{VISPDAT} & $\vi{SA}$ & 325 & 0.98350 & 0.99360 \\
\textbf{VI-FSPDAT} & $\vi{F}$ & 698 & 0.98419 & 0.99216 \\
\textbf{TAY-VISPDAT} & $\vi{Y}$ & 561 & 0.98692 & 0.98853 \\ \hline
\end{tabular}%
}

\caption{Stability of baseline rankings induced by bureaucratic scores. The table lists the number of households for each dataset. The last two columns report the minimum Spearman’s $\rho$ from 10 independent random tie-breaking trials on the SPDAT and RMFS scores. These high values $(\rho > 0.98)$ confirm that these baseline rankings are stable for comparisons with other rankings.}
\label{table:ranking_num_data}
\end{table}
\begin{table}[h]
\resizebox{\columnwidth}{!}{%
\begin{tabular}{|c|c|c|}
\hline
\textbf{\begin{tabular}[c]{@{}c@{}}Assessment\\ Data\end{tabular}} & \textbf{\begin{tabular}[c]{@{}c@{}}Between LLaMA\\ Rankings\end{tabular}} & \textbf{\begin{tabular}[c]{@{}c@{}}Between DeepSeek\\ Rankings\end{tabular}} \\ \hline
\textbf{$\vi{SA}$}                                                  & {\color[HTML]{000000} 0.24692}                                            & {\color[HTML]{000000} 0.47951}                                                  \\
\textbf{$\vi{F}$}                                                & {\color[HTML]{000000} 0.12805}                                            & {\color[HTML]{000000} 0.27571}                                                  \\
\textbf{$\vi{Y}$}                                              & {\color[HTML]{000000} 0.19748}                                            & {\color[HTML]{000000} 0.02719}                                                  \\ \hline
\end{tabular}%
}
\caption{Spearman's $\rho$ between two rankings produced in independent runs by the same model. The generally low correlations indicate that these LLMs produce significantly different rankings when run on the exact same data.}
\label{tab:robustness_spearman}
\end{table}

We now turn to examining LLM judgments in comparison with the judgment of experts on a task involving real-world data.

\paragraph{LLM Internal Consistency} After generating the rankings induced by pairwise comparisons performed by the LLMs, as described in the Methods section, we check for consistency between two different executions of the rank-centrality pipeline. We calculate Spearman's rank correlation between different rankings produced by DS-7B and LL-3-8B and present them in Table \ref{tab:robustness_spearman}.

Rankings produced by LL-3-8B demonstrate weak positive correlations among themselves. DS-7B rankings show moderate positive correlation on the $\vi{SA}$ (single adults) dataset, weak correlation in the $\vi{F}$ (families) dataset, and little to no correlation on the $\vi{Y}$ (youth) dataset. The low degree of correlations presented in Table \ref{tab:robustness_spearman} indicates clearly that LLMs are inconsistent in their vulnerability determinations from questionnaire data commonly used by service providers. 


\begin{table*}[]
\resizebox{\textwidth}{!}{%
\begin{tabular}{lclllc}
\multicolumn{2}{c}{DeepSeek Ranking 1 on $\vi{SA}$} &  & \multicolumn{1}{c}{} & \multicolumn{2}{c}{DeepSeek Ranking 2 on $\vi{SA}$} \\ \cline{1-2} \cline{5-6} 
\multicolumn{1}{|l|}{\textbf{Feature}} & \multicolumn{1}{c|}{\textbf{Coefficient}} &  & \multicolumn{1}{l|}{} & \multicolumn{1}{l|}{\textbf{Feature}} & \multicolumn{1}{c|}{\textbf{Coefficient}} \\ \cline{1-2} \cline{5-6} 
\multicolumn{1}{|l|}{{\color[HTML]{3531FF} Current living arrangement: Motel/Hotel}} & \multicolumn{1}{c|}{{\color[HTML]{FE0000} -0.1498}} &  & \multicolumn{1}{l|}{} & \multicolumn{1}{l|}{{\color[HTML]{000000} Number of past incarcerations $>$ 10}} & \multicolumn{1}{c|}{{\color[HTML]{FE0000} -0.2378}} \\
\multicolumn{1}{|l|}{{\color[HTML]{000000} Current living arrangement: Church}} & \multicolumn{1}{c|}{{\color[HTML]{FE0000} -0.1057}} &  & \multicolumn{1}{l|}{} & \multicolumn{1}{l|}{{\color[HTML]{000000} Current living arrangement: No answer}} & \multicolumn{1}{c|}{{\color[HTML]{FE0000} -0.0400}} \\
\multicolumn{1}{|l|}{{\color[HTML]{000000} Current living arrangement: Family's house}} & \multicolumn{1}{c|}{{\color[HTML]{FE0000} -0.0144}} &  & \multicolumn{1}{l|}{} & \multicolumn{1}{l|}{{\color[HTML]{3531FF} Current living arrangement: Motel/Hotel}} & \multicolumn{1}{c|}{{\color[HTML]{FE0000} -0.0292}} \\
\multicolumn{1}{|l|}{{\color[HTML]{000000} Prescription medication use? No answer}} & \multicolumn{1}{c|}{{\color[HTML]{009901} 0.1475}} &  & \multicolumn{1}{l|}{} & \multicolumn{1}{l|}{{\color[HTML]{000000} Current living arrangement: Job corp}} & \multicolumn{1}{c|}{{\color[HTML]{FE0000} -0.0189}} \\
\multicolumn{1}{|l|}{{\color[HTML]{3531FF} Current living arrangement: Couch-surfing/Shelter}} & \multicolumn{1}{c|}{{\color[HTML]{009901} 0.3555}} &  & \multicolumn{1}{l|}{} & \multicolumn{1}{l|}{{\color[HTML]{3531FF} Current living arrangement: Couch-surfing/Shelter}} & \multicolumn{1}{c|}{{\color[HTML]{009901} 0.3229}} \\ \cline{1-2} \cline{5-6} 
 & \multicolumn{1}{l}{} &  &  &  & \multicolumn{1}{l}{} \\
\multicolumn{2}{c}{DeepSeek Ranking 1 on $\vi{F}$} &  & \multicolumn{1}{c}{} & \multicolumn{2}{c}{DeepSeek Ranking 2 on $\vi{F}$} \\ \cline{1-2} \cline{5-6} 
\multicolumn{1}{|l|}{\textbf{Feature}} & \multicolumn{1}{c|}{\textbf{Coefficient}} &  & \multicolumn{1}{l|}{} & \multicolumn{1}{l|}{\textbf{Feature}} & \multicolumn{1}{c|}{\textbf{Coefficient}} \\ \cline{1-2} \cline{5-6} 
\multicolumn{1}{|l|}{{\color[HTML]{3531FF} Homeless during previous year? No answer}} & \multicolumn{1}{c|}{{\color[HTML]{FE0000} -0.0003}} & {\color[HTML]{000000} } & \multicolumn{1}{l|}{{\color[HTML]{000000} }} & \multicolumn{1}{l|}{{\color[HTML]{000000} Current living arrangement: Relative's house}} & \multicolumn{1}{c|}{{\color[HTML]{FE0000} -0.2136}} \\
\multicolumn{1}{|l|}{{\color[HTML]{000000} Nights spent incarcerated $>$ 3}} & \multicolumn{1}{c|}{{\color[HTML]{009901} 0.0001}} & {\color[HTML]{000000} } & \multicolumn{1}{l|}{{\color[HTML]{000000} }} & \multicolumn{1}{l|}{{\color[HTML]{000000} Current living arrangement: Couch-hopping}} & \multicolumn{1}{c|}{{\color[HTML]{FE0000} -0.0263}} \\
\multicolumn{1}{|l|}{{\color[HTML]{000000} Parents in family $>$ 3}} & \multicolumn{1}{c|}{{\color[HTML]{009901} 0.0002}} & {\color[HTML]{000000} } & \multicolumn{1}{l|}{{\color[HTML]{000000} }} & \multicolumn{1}{l|}{{\color[HTML]{3531FF} Homeless during previous year? No answer}} & \multicolumn{1}{c|}{{\color[HTML]{FE0000} -0.0125}} \\
\multicolumn{1}{|l|}{{\color[HTML]{3531FF} Current living arrangement: No answer}} & \multicolumn{1}{c|}{{\color[HTML]{009901} 0.0002}} & {\color[HTML]{000000} } & \multicolumn{1}{l|}{{\color[HTML]{000000} }} & \multicolumn{1}{l|}{{\color[HTML]{000000} Current living arrangement: Couch-surfing}} & \multicolumn{1}{c|}{{\color[HTML]{009901} 0.0107}} \\
\multicolumn{1}{|l|}{{\color[HTML]{000000} Expected additional children to join after receiving housing: No answer}} & \multicolumn{1}{c|}{{\color[HTML]{009901} 0.0559}} & {\color[HTML]{000000} } & \multicolumn{1}{l|}{{\color[HTML]{000000} }} & \multicolumn{1}{l|}{{\color[HTML]{3531FF} Current living arrangement: No answer}} & \multicolumn{1}{c|}{{\color[HTML]{009901} 0.0220}} \\ \cline{1-2} \cline{5-6} 
{\color[HTML]{000000} } & \multicolumn{1}{l}{{\color[HTML]{000000} }} & {\color[HTML]{000000} } & {\color[HTML]{000000} } & {\color[HTML]{000000} } & \multicolumn{1}{l}{{\color[HTML]{000000} }} \\
\multicolumn{2}{c}{{\color[HTML]{000000} DeepSeek Ranking 1 on $\vi{Y}$}} & {\color[HTML]{000000} } & \multicolumn{1}{c}{{\color[HTML]{000000} }} & \multicolumn{2}{c}{{\color[HTML]{000000} DeepSeek Ranking 2 on $\vi{Y}$}} \\ \cline{1-2} \cline{5-6} 
\multicolumn{1}{|l|}{{\color[HTML]{000000} \textbf{Feature}}} & \multicolumn{1}{c|}{{\color[HTML]{000000} \textbf{Coefficient}}} & {\color[HTML]{000000} } & \multicolumn{1}{l|}{{\color[HTML]{000000} }} & \multicolumn{1}{l|}{{\color[HTML]{000000} \textbf{Feature}}} & \multicolumn{1}{c|}{{\color[HTML]{000000} \textbf{Coefficient}}} \\ \cline{1-2} \cline{5-6} 
\multicolumn{1}{|l|}{{\color[HTML]{000000} Prescription medication use? No answer}} & \multicolumn{1}{c|}{{\color[HTML]{FE0000} -0.0472}} & {\color[HTML]{000000} } & \multicolumn{1}{l|}{{\color[HTML]{000000} }} & \multicolumn{1}{l|}{{\color[HTML]{000000} \# Ambulance rides to hospital $>$ 10}} & \multicolumn{1}{c|}{{\color[HTML]{009901} 0.0371}} \\
\multicolumn{1}{|l|}{{\color[HTML]{3531FF} Alcohol/drug use related eviction? No answer}} & \multicolumn{1}{c|}{{\color[HTML]{FE0000} -0.0472}} & {\color[HTML]{000000} } & \multicolumn{1}{l|}{{\color[HTML]{000000} }} & \multicolumn{1}{l|}{{\color[HTML]{3531FF} Alcohol/drug use impact housing? No answer}} & \multicolumn{1}{c|}{{\color[HTML]{009901} 0.0505}} \\
\multicolumn{1}{|l|}{{\color[HTML]{000000} History of head injury? No answer}} & \multicolumn{1}{c|}{{\color[HTML]{009901} 0.0500}} & {\color[HTML]{000000} } & \multicolumn{1}{l|}{{\color[HTML]{000000} }} & \multicolumn{1}{l|}{{\color[HTML]{000000} Chronic health issues? No answer}} & \multicolumn{1}{c|}{{\color[HTML]{009901} 0.0579}} \\
\multicolumn{1}{|l|}{{\color[HTML]{000000} Marijuana use before age of 12? No answer}} & \multicolumn{1}{c|}{{\color[HTML]{009901} 0.0552}} & {\color[HTML]{000000} } & \multicolumn{1}{l|}{{\color[HTML]{000000} }} & \multicolumn{1}{l|}{{\color[HTML]{000000} $\#$ of interactions with police: 6}} & \multicolumn{1}{c|}{{\color[HTML]{009901} 0.0921}} \\
\multicolumn{1}{|l|}{{\color[HTML]{000000} Used a crisis service? No answer}} & \multicolumn{1}{c|}{{\color[HTML]{009901} 0.1150}} & {\color[HTML]{000000} } & \multicolumn{1}{l|}{{\color[HTML]{000000} }} & \multicolumn{1}{l|}{{\color[HTML]{000000} Current living arrangement: Boyfriend's house}} & \multicolumn{1}{c|}{{\color[HTML]{009901} 0.0927}} \\ \cline{1-2} \cline{5-6} 
\end{tabular}%
}
\caption{Top 5 most influential question-answer pairs for two independent DeepSeek rankings. Similar to Table \ref{fig:top5_featuresllama}, the low overlap and inconsistent polarity of the few shared features (highlighted in blue) demonstrate the model's instability. Questions are shortened for brevity (see Appendix \ref{appendix:ques} \cite{full_paper} for full versions.)}
\label{fig:top5_featuresdeepseek}
\end{table*}

\paragraph{Comparing LLMs with Bureaucratic Scores}
In the US, homelessness services must implement a prioritization system that considers household vulnerability in allocating scarce housing resources, according to federal policies. The St. Louis community has used two different scoring systems, the VI-SPDAT (which has three different forms for three different populations) and the RMFS. How do LLM rankings compare with these well-established systems?\footnote{%
Note that there are many issues researchers and practitioners have identified with such bureaucratic systems, including various kinds of bias, inconsistencies, etc. However, the decision to use these scoring systems at least arises from complex social and political processes that reflect the collective decision of society on how to allocate public benefits and harms.}

\begin{table}[ht]
\resizebox{\columnwidth}{!}{%
\begin{tabular}{|c|c|c|c|c|c|}
\hline
                                        \textbf{\begin{tabular}[c]{@{}c@{}}Assesment \\ Subpopulation\end{tabular}} &
                                         \textbf{\begin{tabular}[c]{@{}c@{}}Ranking \\ Criterion\end{tabular}} & \textbf{\begin{tabular}[c]{@{}c@{}}LLaMA \\ Ranking 1\end{tabular}} & \textbf{\begin{tabular}[c]{@{}c@{}}LLaMA \\ Ranking 2\end{tabular}} & \textbf{\begin{tabular}[c]{@{}c@{}}DeepSeek \\ Ranking 1\end{tabular}} & \textbf{\begin{tabular}[c]{@{}c@{}}DeepSeek \\ Ranking 2\end{tabular}} \\ \hline
                                        & \textbf{VI-SPDAT}                                            & -0.18009                                                            & -0.18640                                                            & {\color[HTML]{000000} 0.12056}                                            & {\color[HTML]{000000} 0.07167}                                            \\
\multirow{-2}{*}{\textbf{$\vi{SA}$}}     & \textbf{RMFS}                                                    & -0.00952                                                            & 0.12469                                                             & {\color[HTML]{000000} 0.16421}                                            & {\color[HTML]{000000} 0.14639}                                            \\ \hline
                                        & \textbf{VI-SPDAT}                                            & -0.06960                                                            & -0.19930                                                            & 0.21722                                                                   & 0.17079                                                                   \\
\multirow{-2}{*}{\textbf{$\vi{F}$}}   & \textbf{RMFS}                                                    & 0.08557                                                             & -0.01491                                                            & 0.12232                                                                   & 0.08018                                                                   \\ \hline
                                        & \textbf{VI-SPDAT}                                            & -0.15286                                                            & -0.16309                                                            & 0.06188                                                                   & 0.04425                                                                   \\
\multirow{-2}{*}{\textbf{$\vi{Y}$}} & \textbf{RMFS}                                                    & -0.01904                                                            & -0.02552                                                            & 0.06974                                                                   & 0.02503                                                                   \\ \hline
\end{tabular}%
}

\caption{Correlation between LLM and bureaucratic rankings. We report the Spearman's $\rho$ between each of the four LLM-generated rankings and the two baseline bureaucratic rankings. The correlations are close to zero, showing that LLM rankings unreliable fail to capture the vulnerability principles embedded in existing systems.}

\label{tab:rank_correlations}
\end{table}

Table \ref{tab:rank_correlations} shows that the rankings by LLMs have near-zero correlation with the VI-SPDAT and RMFS rankings, while LL-3-8B rankings show negative correlation with VI-SPDAT and RMFS. This demonstrates that LLM ranking judgments diverge substantially from those of the vetted bureaucratic ranking systems adopted in coordinated entry systems.

\subsubsection{What Features Drive LLM Decisions? }
To probe deeper into the LLM decision-making criteria, we examine which questions receive higher focus from the models. Each question $q$ in the HMIS questionnaire is accompanied by an answer $\hat{a}$ chosen from a set $A_{q}$ (which could include multiple options written into a catch-all category as well) by the applicant household. For each question answer pair $(q,\hat{a} \in A_{q})$, we generate binary categorical features in the form $(q_{\hat{a}})=1$ and $(q_a)=0$ for all $a\in A_q$ and $a\neq\hat{a}$. Next, we train an ordinal logistic regressor with all thresholds \cite{rennie2005loss} with L2 regularization (coefficient 1) to map these categorical features generated from the questionnaire to the ranks produced by LL-3-8B and DS-7B. We analyze the normalized feature coefficients assigned to each categorical question to better understand which question-answer combinations impact the LLM decisions the most. The LLMs seem to focus on different question-answer pairs in different runs on the same data. As illustration, the top five most important features for LL-3-8B and DS-7B are reported in Tables \ref{fig:top5_featuresllama} and \ref{fig:top5_featuresdeepseek}, respectively. Even when features appear in both rankings, they have inconsistency in the polarity. This shows a struggle to  identify consistent criteria for judging household vulnerability. Interestingly, refusals to answer questionnaire items and unrecorded answers influence both LL-3-8B and DS-7B decisions  on $\vi{F}$ and $\vi{Y}$ subpopulations. Household current living arrangements seem to be a recurring factor impacting the decisions. These may be intuitively aligned with what human perception, however, note that there is considerable inconsistency in which living conditions impact the LLMs most and whether the impact is favorable or adverse to recommending more intensive intervention. \\


\subsubsection{Comparing LLMs and Bureaucratic Scores with Human Decisions}

Although caseworkers are supposed to use bureaucratic scores in determining assignments, street-level bureaucrats inherently can exercise judgment and discretion in deciding who receives services \cite{Lipsky_2010, Pokharel_Das_Fowler_2024, childWelfare2022}. Therefore, it is plausible that LLM judgments match up better with human judgments than bureaucratic scores. In this case, since we only have data on whether or not households received services, we can test how predictive the different rankings are of receipt of intensive services (Rapid Rehousing, Transitional Housing, or Permanent Supportive Housing) by computing the area under the ROC curve for predicting receipt of one of these services for each of the rankings.

\begin{table}[ht]
\resizebox{\columnwidth}{!}{%
\begin{tabular}{|l|l|l|l|l|l|l|l|}
\hline
\textbf{\begin{tabular}[c]{@{}l@{}}Assessment \\ Data\end{tabular}} & \textbf{\begin{tabular}[c]{@{}l@{}}\# Positive\\ Samples\end{tabular}} & \textbf{\begin{tabular}[c]{@{}l@{}}VI-SPDAT\\ Ranking\end{tabular}} & \textbf{\begin{tabular}[c]{@{}l@{}}RMFS\\ Ranking\end{tabular}} & \textbf{\begin{tabular}[c]{@{}l@{}}LLaMA \\ Ranking 1\end{tabular}} & \textbf{\begin{tabular}[c]{@{}l@{}}LLaMA \\ Ranking 1\end{tabular}} & \textbf{\begin{tabular}[c]{@{}l@{}}DeepSeek-R1 \\ Ranking 1\end{tabular}} & \textbf{\begin{tabular}[c]{@{}l@{}}DeepSeek-R1 \\ Ranking 2\end{tabular}} \\ \hline
$\vi{SA}$ & 68 & 0.68511 & 0.6042 & 0.61667 & 0.62434 & 0.51957 & 0.51247 \\
$\vi{F}$ & 116 & 0.53022 & 0.5253 & 0.51518 & 0.5374 & 0.51271 & 0.54361 \\
$\vi{Y}$ & 103 & 0.50674 & 0.56923 & 0.51015 & 0.53258 & 0.54498 & 0.50854 \\ \hline
\end{tabular}%
}

\caption{Predictive validity of rankings for service allocation. The table reports the ROC AUC for each ranking's ability to predict the receipt of an intensive housing intervention, with the number of positive samples (recipients) noted. All rankings are weak predictors, and LLM-generated rankings offer no improvement over existing bureaucratic tools in forecasting real-world decisions on prioritization. }
\label{tab:roc_intensive_match}
\end{table}



Table \ref{tab:roc_intensive_match} presents the revealing results. The bureaucratic scores are only predictive for single adults; LLaMA is also somewhat predictive in this case, but less so than the VI-SPDAT scores. Meanwhile, for families and youth, neither the bureaucratic scores nor the LLM rankings predict actual caseworker decisions. There are many factors determining if a household receives services, especially the availability of appropriate housing at the time of assessment. We should not expect very high AUC values, but the \emph{differences} are illuminating. In short, LLM rankings show virtually no correlation with well-known scoring systems and are less predictive of the actual decision-making of human experts.  

\section{Conclusion}

In this work, we examine the viability of deploying off-the-shelf LLMs as de facto street-level bureaucrats in the context of homelessness resource allocation (``vibe prioritization''). Through two complementary experimental tasks -- pairwise comparisons and global ranking reconstructed via Rank Centrality -- we systematically evaluate (1) the internal consistency of LLM judgments, (2) their agreement with established bureaucratic scoring systems, and (3) their alignment with actual caseworker decisions.

Our pairwise comparison experiments (Section \ref{subsec:pariwise_results}) demonstrate that, when presented with household profiles alone, LLMs exhibit variability reminiscent of non-expert human subjects: different LLMs are different in whether they naturally take vulnerability-oriented decisions or outcome-oriented ones, but relatively consistent within themselves \emph{as long as they receive explicit risk information} (computed by an external conventional machine learning model) 
(Figures \ref{fig:model_level_results} and   \ref{fig:block_level_results}). They are, however, inconsistent in decision-making when not provided explicit risk information.

The ranking task (Section \ref{subsec:rankings_results}) reveals  serious limitations. 
Neither LL-3-8B nor DS-7B produces stable vulnerability orderings: Spearman correlations between independent runs range from near zero to moderate (Table \ref{tab:robustness_spearman}). This means that two executions of the same model can generate markedly different rankings for the same set of households, calling into question the robustness and reliability required for high-stakes decision-making. Furthermore, LLM-generated rankings show negligible, and even negative, correlation with bureaucratic scoring systems (Table \ref{tab:rank_correlations}), as well as being less predictive of actual homelessness service allocations (Table \ref{tab:roc_intensive_match}). Taken together, the results signal that LLMs, without domain-specific adaptation, fail to replicate either formalized points-based systems or the nuanced discretion of experienced caseworkers.

These findings raise important concerns for policymakers and practitioners considering integrating LLMs into high-stakes social service workflows. First, the pronounced inconsistency suggests that automated judgments may vary dramatically depending on incidental factors -- model choice, prompt phrasing, or inference seed -- undermining fairness and transparency. Second, the disconnect from established vulnerability metrics risks both inefficient resource deployment and erosion of community-driven prioritization principles. This misalignment indicates that LLMs may rely on superficial patterns rather than the policy-driven indicators of vulnerability, potentially exacerbating service gaps and deepening inequities. They are also not consistent with any of the prioritization principles that have been developed across many different domains of local justice.

Finally, the limited ability of models to predict real allocations indicates insufficient alignment with the tacit expertise of street-level bureaucrats. As they are, without careful consideration or thorough testing, off-the-shelf LLMs fail to capture the local justice principles and context-sensitive discretion that street-level bureaucrats embed into their decisions \cite{Lipsky_2010, elster1992local}, emphasizing that LLM-driven approaches as of now cannot supplant embodied professional judgment.


This study underscores the necessity of rigorous context-grounded evaluation before automating public resource allocation decisions. Future work must investigate whether practices like fine-tuning on localized caseworker data, integrating multi-modal client information, or embedding human-in-the-loop safeguards can enhance LLM reliability. Equally crucial is the need for ongoing engagement with service providers and stakeholders to ensure that algorithmic systems respect norms of justice and preserve the discretionary judgment essential to public service. Importantly, any AI augmentation in this domain must reflect the nuanced trade-offs and moral frameworks that have evolved through community engagement and political processes.

In conclusion, while LLMs offer promising avenues for augmenting decision support, the evidence from our study cautions against their wholesale replacement of street-level bureaucrats in homelessness services. Realizing the potential of AI in this domain requires carefully calibrated hybrid systems, where machine recommendations are transparent, consistently calibrated, and subject to human oversight, rather than unmediated reliance on present-day language models.

\section*{Acknowledgments}
We are grateful for support from NSF Award 2533162. We also thank the various community partners who helped conceptualize the challenges facing the delivery of homeless services, as well as their ongoing efforts to support local families.

\bibliography{references}

\newpage 
\onecolumn
\appendix

\section{Table Results with Confidence Intervals}
\label{Appendix:All_Tables}

\begin{table*}[ht]      
  \centering
  \caption{Experimental Results with 95\% Confidence Intervals. These tables provide comprehensive statistical details for the results presented in the main body, including dataset sizes, baseline score stability, LLM reliability, and predictive performance.}
  \label{tbl:all_results_CI}

  \begin{subtable}[t]{\textwidth}
    \centering
    \resizebox{\textwidth}{!}{%
    \begin{tabular}{|c|c|c|cc|cc|}
      \hline
      \multirow{2}{*}{\textbf{Assessment Type}} & \multirow{2}{*}{\textbf{Dataset Name}} & \multirow{2}{*}{\textbf{Num Samples}} & \multicolumn{2}{c|}{\textbf{SPDAT}} & \multicolumn{2}{c|}{\textbf{RMFS}} \\ \cline{4-7} 
      & & & \multicolumn{1}{c|}{\textbf{Min $\rho$}} & \textbf{95\% CI} & \multicolumn{1}{c|}{\textbf{Min $\rho$}} & \textbf{95\% CI} \\ \hline
      \textbf{VISPDAT} & $\vi{SA}$ & 325 & \multicolumn{1}{c|}{0.98350} & (0.97852, 0.98733) & \multicolumn{1}{c|}{0.99360} & (0.99165, 0.99510) \\
      \textbf{VIFSPDAT} & $\vi{F}$ & 698 & \multicolumn{1}{c|}{0.98419} & (0.98107, 0.98679) & \multicolumn{1}{c|}{0.99216} & (0.99060, 0.99346) \\
      \textbf{TAY} & $\vi{Y}$ & 561 & \multicolumn{1}{c|}{0.98692} & (0.98401, 0.98930) & \multicolumn{1}{c|}{0.98853} & (0.98598, 0.99063) \\ \hline
    \end{tabular}%
    }
    \caption{Dataset summary and stability of bureaucratic rankings (Table \ref{table:ranking_num_data}) after tie-breaking. Minimum Spearman's $\rho$ and 95\% CIs \cite{Bonett_Wright_2000} are reported from 10 random tie-breaking trials.}
    \label{labeltable:ranking_num_data_CI}
  \end{subtable}

  \vspace{1.5em} 

  \begin{subtable}[t]{\textwidth}
    \centering
    \resizebox{0.65\textwidth}{!}{
    \begin{tabular}{|c|cc|cc|}
      \hline
      \multirow{2}{*}{\textbf{\begin{tabular}[c]{@{}c@{}}Assessment \\ Data\end{tabular}}} & \multicolumn{2}{c|}{\textbf{Between LLaMA Rankings}} & \multicolumn{2}{c|}{\textbf{Between DeepSeek Rankings}} \\ \cline{2-5} 
      & \multicolumn{1}{c|}{\textbf{$\rho$}} & \textbf{95\% CI} & \multicolumn{1}{c|}{\textbf{$\rho$}} & \textbf{95\% CI} \\ \hline
      $\vi{SA}$ & \multicolumn{1}{c|}{0.24692} & (0.14032, 0.34786) & \multicolumn{1}{c|}{0.47951} & (0.38594, 0.56332) \\
      $\vi{F}$ & \multicolumn{1}{c|}{0.12805} & (0.05405, 0.20064) & \multicolumn{1}{c|}{0.27571} & (0.20437, 0.34414) \\
      $\vi{Y}$ & \multicolumn{1}{c|}{0.19748} & (0.11581, 0.27650) & \multicolumn{1}{c|}{0.02719} & (-0.05573, 0.10974) \\ \hline
    \end{tabular}%
    }
    \caption{Run-to-run reliability of LLM rankings (Table \ref{tab:robustness_spearman}). Spearman's $\rho$ and 95\% CIs \cite{Bonett_Wright_2000} are shown between two independent runs of the same model.}
    \label{labeltable:robustness_spearman_CI}
  \end{subtable}

  \vspace{1.5em}

  
  \begin{subtable}[t]{\textwidth}
    \centering
    \resizebox{\textwidth}{!}{%
    \begin{tabular}{|c|c|cc|cc|cc|cc|}
      \hline
      \multirow{2}{*}{\textbf{\begin{tabular}[c]{@{}c@{}}Assessment\\  Subpopulation\end{tabular}}} & \multirow{2}{*}{\textbf{\begin{tabular}[c]{@{}c@{}}Ranking \\ Criterion\end{tabular}}} & \multicolumn{2}{c|}{\textbf{\begin{tabular}[c]{@{}c@{}}LLaMA \\ Ranking 1\end{tabular}}} & \multicolumn{2}{c|}{\textbf{\begin{tabular}[c]{@{}c@{}}LLaMA \\ Ranking 2\end{tabular}}} & \multicolumn{2}{c|}{\textbf{\begin{tabular}[c]{@{}c@{}}DeepSeek \\ Ranking 1\end{tabular}}} & \multicolumn{2}{c|}{\textbf{\begin{tabular}[c]{@{}c@{}}DeepSeek \\ Ranking 2\end{tabular}}} \\ \cline{3-10} 
      & & \multicolumn{1}{c|}{\textbf{$\rho$}} & \textbf{95\% CI} & \multicolumn{1}{c|}{\textbf{$\rho$}} & \textbf{95\% CI} & \multicolumn{1}{c|}{\textbf{$\rho$}} & \textbf{95\% CI} & \multicolumn{1}{c|}{\textbf{$\rho$}} & \textbf{95\% CI} \\ \hline
      \multirow{2}{*}{$\vi{SA}$} & \textbf{VI-SPDAT} & \multicolumn{1}{c|}{-0.18009} & (-0.28414, -0.07185) & \multicolumn{1}{c|}{-0.18640} & (-0.29019, -0.07828) & \multicolumn{1}{c|}{0.12056} & (0.01152, 0.22676) & \multicolumn{1}{c|}{0.07167} & (-0.03756, 0.17920) \\
      & \textbf{RMFS} & \multicolumn{1}{c|}{-0.00952} & (-0.11819, 0.09938) & \multicolumn{1}{c|}{0.12469} & (0.01569, 0.23076) & \multicolumn{1}{c|}{0.16421} & (0.05570, 0.26889) & \multicolumn{1}{c|}{0.14639} & (0.03762, 0.25172) \\ \hline
      \multirow{2}{*}{$\vi{F}$} & \textbf{VI-SPDAT} & \multicolumn{1}{c|}{-0.06960} & (-0.14316, 0.00472) & \multicolumn{1}{c|}{-0.19930} & (-0.27021, -0.12625) & \multicolumn{1}{c|}{0.21722} & (0.14450, 0.28761) & \multicolumn{1}{c|}{0.17079} & (0.09728, 0.24244) \\
      & \textbf{RMFS} & \multicolumn{1}{c|}{0.08557} & (0.01130, 0.15890) & \multicolumn{1}{c|}{-0.01491} & (-0.08902, 0.05937) & \multicolumn{1}{c|}{0.12232} & (0.04827, 0.19503) & \multicolumn{1}{c|}{0.08018} & (0.00589, 0.15359) \\ \hline
      \multirow{2}{*}{$\vi{Y}$} & \textbf{VI-SPDAT} & \multicolumn{1}{c|}{-0.15286} & (-0.23316, -0.07050) & \multicolumn{1}{c|}{-0.16309} & (-0.24311, -0.08086) & \multicolumn{1}{c|}{0.06188} & (-0.02109, 0.14400) & \multicolumn{1}{c|}{0.04425} & (-0.03871, 0.12661) \\
      & \textbf{RMFS} & \multicolumn{1}{c|}{-0.01904} & (-0.10167, 0.06385) & \multicolumn{1}{c|}{-0.02552} & (-0.10809, 0.05740) & \multicolumn{1}{c|}{0.06974} & (-0.01322, 0.15174) & \multicolumn{1}{c|}{0.02503} & (-0.05788, 0.10760) \\ \hline
    \end{tabular}%
    }
    \caption{Agreement between LLM and bureaucratic rankings (Table \ref{tab:rank_correlations}). Spearman's $\rho$ and 95\% CIs \cite{Bonett_Wright_2000} between each LLM ranking and the baseline scores.}
    \label{labeltable:rank_correlations_CI}
  \end{subtable}

  \vspace{1.5em}

    \begin{subtable}[t]{\textwidth}
    \centering
    \resizebox{\textwidth}{!}{%
    \begin{tabular}{|c|c|cc|cc|cc|cc|cc|cc|}
      \hline
      \multirow{2}{*}{\textbf{\begin{tabular}[c]{@{}c@{}}Assessment \\ Data\end{tabular}}} & \multirow{2}{*}{\textbf{\begin{tabular}[c]{@{}c@{}}\# Positive\\ Samples\end{tabular}}} & \multicolumn{2}{c|}{\textbf{\begin{tabular}[c]{@{}c@{}}VI-SPDAT \\ Ranking\end{tabular}}} & \multicolumn{2}{c|}{\textbf{\begin{tabular}[c]{@{}c@{}}RMFS \\ Ranking\end{tabular}}} & \multicolumn{2}{c|}{\textbf{\begin{tabular}[c]{@{}c@{}}LLaMA \\ Ranking 1\end{tabular}}} & \multicolumn{2}{c|}{\textbf{\begin{tabular}[c]{@{}c@{}}LLaMA \\ Ranking 2\end{tabular}}} & \multicolumn{2}{c|}{\textbf{\begin{tabular}[c]{@{}c@{}}DeepSeek \\ Ranking 1\end{tabular}}} & \multicolumn{2}{c|}{\textbf{\begin{tabular}[c]{@{}c@{}}DeepSeek \\ Ranking 2\end{tabular}}} \\ \cline{3-14} 
      & & \multicolumn{1}{c|}{\textbf{AUC}} & \textbf{95\% CI} & \multicolumn{1}{c|}{\textbf{AUC}} & \textbf{95\% CI} & \multicolumn{1}{c|}{\textbf{AUC}} & \textbf{95\% CI} & \multicolumn{1}{c|}{\textbf{AUC}} & \textbf{95\% CI} & \multicolumn{1}{c|}{\textbf{AUC}} & \textbf{95\% CI} & \multicolumn{1}{c|}{\textbf{AUC}} & \textbf{95\% CI} \\ \hline
      $\vi{SA}$ & 68 & \multicolumn{1}{c|}{0.69404} & (0.62449, 0.76359) & \multicolumn{1}{c|}{0.59791} & (0.52841, 0.6674) & \multicolumn{1}{c|}{0.61667} & (0.54023, 0.69312) & \multicolumn{1}{c|}{0.62434} & (0.54843, 0.70026) & \multicolumn{1}{c|}{0.51957} & (0.44263, 0.59651) & \multicolumn{1}{c|}{0.51247} & (0.43261, 0.59234) \\
      $\vi{F}$ & 116 & \multicolumn{1}{c|}{0.52804} & (0.47414, 0.58193) & \multicolumn{1}{c|}{0.52422} & (0.46678, 0.58165) & \multicolumn{1}{c|}{0.51518} & (0.45844, 0.57192) & \multicolumn{1}{c|}{0.53740} & (0.48256, 0.59224) & \multicolumn{1}{c|}{0.51271} & (0.45238, 0.57304) & \multicolumn{1}{c|}{0.54361} & (0.48757, 0.59965) \\
      $\vi{Y}$ & 103 & \multicolumn{1}{c|}{0.51304} & (0.45453, 0.57154) & \multicolumn{1}{c|}{0.57135} & (0.51352, 0.62918) & \multicolumn{1}{c|}{0.51015} & (0.44594, 0.57437) & \multicolumn{1}{c|}{0.53258} & (0.46859, 0.59658) & \multicolumn{1}{c|}{0.54498} & (0.48796, 0.60201) & \multicolumn{1}{c|}{0.50854} & (0.44528, 0.57180) \\ \hline
    \end{tabular}%
    }
    \caption{Predictive validity of rankings for service allocation (Table \ref{tab:roc_intensive_match}). ROC AUC scores and 95\% CIs \cite{DeLong_DeLong_Clarke-Pearson_1988} for predicting the receipt of an intensive housing intervention.}
    \label{labeltable:roc_intensive_match_CI}
  \end{subtable}
  \label{tab:all_results}
\end{table*}
\section{Questionnaires used in VI-SPDAT system to score vulnurablity}
\label{appendix:ques}
\subsection{VI-SPDAT}
\label{appen:visp}
\begin{enumerate}

    \item Are you currently able to care for your basic needs, such as bathing, changing clothes, using the restroom, obtaining food, and accessing clean water?
    \item Are you not taking any medications that a doctor has prescribed for you?
    \item Are you taking prescribed painkillers incorrectly or selling them instead of using them as directed?
    \item Do you currently have legal issues that might result in incarceration, fines, or difficulties in renting housing?
    \item Do you engage in risky behaviors, such as exchanging sex for money, running drugs, having unprotected sex with strangers, sharing needles, or similar activities?
    \item Do you have a learning disability, developmental disability, or any other impairment?
    \item Do you have a mental health issue or concern?
    \item Do you have any mental health or cognitive issues that make it difficult to live independently?
    \item Do you have any physical disabilities that limit the type of housing you can access or make it difficult to live independently?
    \item Do you have planned activities—aside from mere survival—that make you feel happy and fulfilled?
    \item Do you receive income from the government, a pension, an inheritance, informal work, or a regular job?
    \item Do you suffer from any chronic health issues involving your liver, kidneys, stomach, lungs, or heart?
    \item Does anyone force or trick you into doing things against your will?
    \item For female respondents only: Are you currently pregnant?
    \item Has your alcohol or drug use resulted in you being kicked out of an apartment or shelter program in the past?
    \item Has your current period of homelessness been caused by experiencing emotional, physical, psychological, sexual, or other trauma? (Please answer YES or NO)
    \item Have you been attacked or beaten up since you became homeless?
    \item Have you been hospitalized as an inpatient?
    \item Have you ever had to leave your apartment, shelter program, or other living arrangement because of physical health issues?
    \item Have you experienced a head injury in the past?
    \item Have you received health care at an emergency department or room?
    \item Have you spent one or more nights in a holding cell, jail, or prison, regardless of the duration?
    \item Have you taken an ambulance to the hospital?
    \item Have you talked to the police because you witnessed a crime, were a victim or suspect, or were told to move along?
    \item Have you used a crisis service, such as those for sexual assault, mental health, family/intimate violence, distress, or suicide prevention?
    \item How long has it been since you lived in permanent, stable housing?
    \item If space were available in a program that specifically assists people living with HIV or AIDS, would you be interested?
    \item In the last three years, how many times have you experienced homelessness?
    \item In the last year, have you threatened or attempted to harm yourself or someone else?
    \item Is there any person or entity (for example, a past landlord, business, bookie, dealer, or government group like the IRS) that believes you owe them money?
    \item Is your current homelessness caused by a relationship breakdown, an unhealthy or abusive relationship, or actions by family or friends leading to eviction?
    \item When you are sick or not feeling well, do you avoid seeking help?
    \item Where do you sleep most frequently? (choose one)
    \begin{itemize}
        \item If you selected `Other', please specify further details.
    \end{itemize}
    \item Will alcohol or drug use make it difficult for you to maintain or afford housing?

\end{enumerate}

\subsection{VI-F-SPDAT}
\label{appens:vifsp}
\begin{enumerate}

    \item Has any child in the family experienced abuse or trauma in the last 180 days?
    \item Stayed one or more nights in a holding cell, jail, or prison, whether that was a short-term stay like the drunk tank, a longer stay for a more serious offense, or anything in between?
    \item Are any prescribed painkillers not taken as directed or being sold instead of used?
    \item Are there any children in your family aged 11 or younger?
    \item Are there any children in your family aged 6 or younger?
    \item Are there any family legal issues currently in or needing court resolution that might impact your housing or household composition?
    \item Do children aged 12 or younger spend 2 or more hours per day in activities?
    \item Do children aged 13 or older spend 3 or more hours per day in activities?
    \item Do you anticipate any additional adults or children joining your household within the first 180 days of being housed?
    \item Do your older children spend 2 or more hours per day helping their younger siblings with tasks such as preparing for school, homework, dinner, or bathing?
    \item Does any single member of your household have a medical condition, mental health concern, and issues with substance use?
    \item Does anyone in your family currently have legal issues that might lead to incarceration, fines, or difficulties in renting housing?
    \item Does anyone in your family force or trick you into doing things against your will?
    \item Does anyone in your family have a learning disability, developmental disability, or any other impairment?
    \item Does anyone in your family have a mental health issue or concern?
    \item Does anyone in your family have mental health or cognitive issues that make independent living challenging?
    \item Does anyone in your family have physical disabilities that limit the type of housing accessible or make independent living challenging?
    \item Does anyone in your family suffer from chronic health issues involving their liver, kidneys, stomach, lungs, or heart?
    \item Does every member of your family have planned activities—beyond just survival—that bring happiness and fulfillment?
    \item Does your family engage in risky behaviors such as exchanging sex for money, running drugs, having unprotected sex, sharing needles, or similar activities?
    \item Does your family have two or more planned activities each week, such as outings, library visits, family gatherings, or movie nights?
    \item Does your family receive income from the government, a pension, inheritance, informal work, or a regular job?
    \item Has alcohol or drug use by anyone in your family resulted in being evicted from your housing or shelter program?
    \item Has any family member been hospitalized as an inpatient?
    \item Has any family member received care at an emergency department or room?
    \item Has any family member taken an ambulance to the hospital?
    \item Has anyone in your family been attacked or beaten up since becoming homeless?
    \item Has anyone in your family experienced a head injury in the past?
    \item Has your family composition changed in the last 180 days due to factors such as divorce, children rejoining, someone leaving for military service or incarceration, or a relative moving in?
    \item Has your family ever had to leave an apartment, shelter, or other accommodation due to physical health issues?
    \item Has your family spoken to the police because they witnessed a crime, were victims, or were suspects, or because the police advised them to move along?
    \item Has your family used a crisis service such as those for sexual assault, mental health, family/intimate violence, distress, or suicide prevention?
    \item Has your family's current homelessness been caused by experiencing emotional, physical, psychological, sexual, or other trauma? (Please answer YES or NO)
    \item Have any children been removed from your family by a child protection service in the last 180 days?
    \item How long has it been since you and your family lived in permanent, stable housing?
    \item How many children under the age of 18 are currently with you?
    \item How many children under the age of 18 are not currently with your family but are expected to join you when you get housed?
    \item How many parents are in your family?
    \item If space were available in a program for people living with HIV or AIDS, would your family be interested?
    \item If there are school-aged children, do they attend school most weeks?
    \item If your household includes a female, is any member currently pregnant?
    \item In the last 180 days, have any children lived with family or friends due to your housing situation or homelessness?
    \item In the last three years, how many times has your family experienced homelessness?
    \item In the last year, has anyone in your family threatened or attempted to harm themselves or others?
    \item Is every member of your family capable of taking care of basic needs such as bathing, changing clothes, using the restroom, obtaining food, and accessing clean water?
    \item Is the head of your household 60 years of age or older?
    \item Is there any medication that a doctor prescribed for you or your family that is not being taken?
    \item Is there any person or entity (e.g., past landlord, business, bookie, dealer, government group like the IRS) that believes your family owes them money?
    \item Is your family's current homelessness caused by a relationship breakdown, an unhealthy or abusive relationship, or interventions by family or friends?
    \item Please provide a list of children's names and ages.
    \item When a family member is sick or unwell, does your family avoid seeking medical help?
    \item Where do you and your family sleep most frequently? (choose one)
    \begin{itemize}
        \item If you selected `Other', please specify further details.
    \end{itemize}
    \item Will alcohol or drug use make it difficult for your family to maintain or afford housing?
    
\end{enumerate}

\subsection{TAY-VI-SPDAT}
\label{appen:tay}
\begin{enumerate}

    \item Are you currently able to care for your basic needs (such as bathing, changing clothes, using a restroom, obtaining food, and accessing clean water)?
    \item Are you not taking any medications that a doctor prescribed for you?
    \item Are you taking prescribed painkillers incorrectly or selling them instead of using them as directed?
    \item Did conflicts regarding gender identity or sexual orientation contribute to your homelessness?
    \item Did differences in religious or cultural beliefs with your parents, guardians, or caregivers lead to your homelessness?
    \item Did you become homeless because you ran away from your family home, a group home, or a foster home?
    \item Do you currently have legal issues that might lead to incarceration, fines, or difficulties renting housing?
    \item Do you engage in risky behaviors (such as exchanging sex for money, food, drugs, or a place to stay; running drugs; having unprotected sex with strangers; sharing needles; etc.)?
    \item Do you have a learning disability, developmental disability, or any other impairment?
    \item Do you have a mental health issue or concern?
    \item Do you have any mental health or cognitive issues that hinder your ability to live independently?
    \item Do you have physical disabilities that limit your housing options or make living independently difficult?
    \item Do you have planned activities—beyond mere survival—that make you feel happy and fulfilled?
    \item Do you receive income from the government, an inheritance, an allowance, informal work, or a regular job?
    \item Do you suffer from chronic health issues involving your liver, kidneys, stomach, lungs, or heart?
    \item Does anyone force or trick you into doing things against your will?
    \item Has your alcohol or drug use resulted in eviction from your housing or shelter program?
    \item Have you been attacked or beaten up since you became homeless?
    \item Have you been hospitalized as an inpatient?
    \item Have you ever been pregnant, or have you ever gotten someone pregnant?
    \item Have you ever had to leave your living situation due to physical health issues?
    \item Have you experienced a head injury in the past?
    \item Have you received care at an emergency department or room?
    \item Have you spent one or more nights in a holding cell, jail, prison, or juvenile detention, regardless of the duration?
    \item Have you taken an ambulance to the hospital?
    \item Have you talked to the police because you witnessed a crime, were a victim or suspect, or were advised to move along?
    \item Have you used a crisis service (for sexual assault, mental health, family/intimate violence, distress, or suicide prevention)?
    \item How long has it been since you lived in permanent, stable housing?
    \item If space were available in a program for people living with HIV or AIDS, would you be interested?
    \item If you have tried marijuana, did you first try it at age 12 or younger?
    \item In the last three years, how many times have you experienced homelessness?
    \item In the last year, have you threatened or attempted to harm yourself or others?
    \item Is there any person or entity (e.g., a past landlord, business, bookie, dealer, or the IRS) that believes you owe money?
    \item Was your homelessness a result of violence among family members at home?
    \item Was your homelessness caused by an unhealthy or abusive relationship?
    \item Was your homelessness caused by your family or friends?
    \item Were you ever incarcerated before the age of 18?
    \item When you are sick or unwell, do you avoid seeking medical help?
    \item Where do you sleep most frequently? (choose one)
    \begin{itemize}
        \item If you selected `Other', please specify further details.
    \end{itemize}
    \item Will alcohol or drug use make it difficult for you to maintain or afford housing?
\end{enumerate}

\section{Questionnaires used in RMFS system to score vulnurablity}
\label{appendix:ques_rmfs}

\begin{enumerate}
\item Age
\item Some housing service agencies provide specialized services for households that include pregnant women. Is anyone in the household pregnant?
\item  Are you or any other adults in the household responsible for a child living outside of the current household?
\item Chronic Kidney Disease (undergoing dialysis)
\item Chronic Lung Disease
\item Diabetes
\item Does anyone in the household have a poor credit history, making it hard to rent a home?
\item Does your household have a support network that can help you with your kids or anything else if it comes up?
\item Have you or any other adult in the household experienced homelessness or unstable housing in your life?
\item Have you or anyone in your household been affected by the justice system, making it difficult to rent a home?
\item Have you or anyone in your household been in the Foster Care system?
\item Have you or anyone in your household ever been evicted?
\item Have you or anyone in your household experienced trauma related to law enforcement encounters?
\item How long has it been since you and your household lived in stable permanent housing?
\item If medication is prescribed for any of the conditions above, is that medication taken as prescribed?
\item Immunocompromised?
\item In any of the times above, was an ambulance taken to the hospital?
\item In any of the times above, was there a hospital stay for more than 24 hours?
\item In the past six months, how many times have you (and/or your household) been to an emergency room to receive healthcare services?
\item Is a doctor being seen for any of the above conditions?
\item Liver Disease
\item Serious Heart Condition
\item  Severe Asthma
\item Severe Obesity
\item Where did you sleep most frequently during this period of homelessness?

\end{enumerate}

\end{document}